\documentclass[preprint,aps]{revtex4}
\usepackage{amssymb, amsmath}
\usepackage[final]{graphicx}
\usepackage[latin1]{inputenc}
\usepackage{color}

\begin{document}
\preprint{ITP-UU-07/9, SPIN-07/9}
\title{\Large\bf Vacuum properties of nonsymmetric gravity in de Sitter space}

\author{Tomas Janssen}
\email[]{T.Janssen2@phys.uu.nl}
\author{Tomislav Prokopec}
\email[]{T.Prokopec@phys.uu.nl}

\affiliation{Institute for Theoretical Physics (ITP) \& Spinoza Institute,
             Utrecht University, Leuvenlaan 4, Postbus 80.195,
              3508 TD Utrecht, The Netherlands}

\begin{abstract}
  We consider quantum effects of a massive antisymmetric tensor field
on the dynamics of de Sitter space-time. Our starting point is the
most general, stable, linearized Lagrangian arising in
nonsymmetric gravitational theories (NGTs), where part of the
antisymmetric field mass is generated by the cosmological term. We
construct a renormalization group (RG) improved effective action
by integrating out one loop vacuum fluctuations of the
antisymmetric tensor field and show that, in the limit when the RG
scale goes to zero, the Hubble parameter -- and thus the effective
cosmological constant -- relaxes rapidly to zero. We thus conclude
that quantum loop effects in de Sitter space can dramatically
change the infrared sector of the on-shell gravity, making the
expansion rate insensitive to the original (bare) cosmological
constant.
\end{abstract}



 \maketitle

\section{Introduction}
In modern quantum field theories, the vacuum is never really
empty. A clear example of this is the harmonic oscillator
with fundamental frequency $\omega$, whose
ground state has an energy $E=\hbar\omega/2$. This nonzero
'zero point' energy is interpreted as
the energy that is always present, even when the oscillator is
not excited. One could see this as a manifestation of Heisenberg's
uncertainty principle: the oscillator is never completely at
rest~\cite{Weinberg:1988cp,Nobbenhuis:2006yf}.
\\
The zero point energy of the quantum harmonic oscillator is an
example of a very general phenomenon in quantum (field) theories.
The quantum vacuum gets enhanced by loop Feynman diagrams. In
quantum field theory these diagrams represent a shift in the
potential energy and such a shift is, like in classical mechanics,
in general unobservable. Therefore often the normal ordering
prescription is used, which removes the zero point energy.
However, there are special cases, like the Casimir effect, where
the vacuum energy does become important. The experimental
verification~\cite{Lamoreaux:1996wh} of the Casimir
effect~\cite{Casimir:1948dh} however, may not prove the existence
of the vacuum energy~\cite{Jaffe:2005vp}.
\\
 \subsection{The cosmological constant problem}

 The problems with vacuum energy start to arise as soon as we start
to consider general relativity (GR). Whereas in quantum theories,
energies are usually observable as the difference between some
excitation of the Hamiltonian and the ground state, this is not
the case in GR. In GR curvature is sourced by the energy momentum
tensor $T_{\mu\nu}$ according to Einstein's equations
\begin{equation}\label{einstein eq}
    R_{\mu\nu}-\frac{1}{2}R g_{\mu\nu}-\Lambda_g
    g_{\mu\nu}=-\frac{1}{2Q}T_{\mu\nu},
\end{equation}
where $R$ and $R_{\mu\nu}$ are the Ricci scalar and tensor
respectively, $g_{\mu\nu}$ the metric tensor, $Q$ is related to
Newtons constant $G_N$ as  $Q\equiv(16\pi G_N)^{-1}$ and
$\Lambda_g$ is a possible geometrical cosmological constant.
$T_{\mu\nu}$ is sourced by any form of energy, including the
vacuum energy and in fact Lorentz invariance implies that
\begin{equation}
    T^{\text{vac}}_{\mu\nu}=-\langle\rho\rangle g_{\mu\nu},
\end{equation}
where $\langle\rho\rangle$ is the energy density associated with
the zero point energy. It is thus clear from (\ref{einstein eq})
that the zero point energy sources gravity in exactly the same way
as a positive cosmological constant would. In fact we could write
\begin{equation}
    \Lambda_{\rm c}=\Lambda_{\rho}+\Lambda_g,
\end{equation}
where $\Lambda_{\rm c}$ is the total, classical cosmological
constant. It is a measurable quantity and current
observations~\cite{Perlmutter:1998np,Riess:1998cb,Astier:2005qq,Riess:2004nr,Spergel:2006hy}
indicate that
\begin{equation}
    \Lambda_{\rm c}\leq 3H_0^2\Omega_{\rm DE}
              \simeq 5\times 10^{-84}\, (\text{GeV})^2
\,,
\label{Lambda:obs}
\end{equation}
where $H_0\simeq 1.5\times 10^{-42}~{\rm GeV}$ denotes the Hubble parameter
today and $\Omega_{\rm DE} \simeq 0.73$ is the dark energy density
in the units of the critical density.
However, standard calculations in quantum theory tell us that
\begin{equation}\label{QFT lambda}
    \Lambda_{\rho}\sim \frac{M_X^4}{m_p^2},
\end{equation}
where $m_p=1.2\times 10^{19}\,\text{GeV}$ is the Planck mass and
$M_X$ is the maximum mass scale which contributes to the vacuum
energy. In the most optimistic case, this could be the QCD scale
($\sim 200\,\text{MeV}$), but more realistically it may be as high
as $M_{\rm GUT}\sim 10^{16}~{\rm GeV}$ or even $m_p$. The
cosmological constant problem can be formulated as the huge
discrepancy between the observed value $\Lambda_{\rm c}$ and
$\Lambda_{\rho}$.
 Even if $M_X$ is at the QCD scale
the difference is 40 orders of magnitude, while if $M_X$ is at the
Planck scale, the difference is an astonishing 122 orders of
magnitude. Of course one could tune $\Lambda_g$ to fit the
observations. This however seems very unnatural, in particular
when one takes account of phase transitions in the early Universe
during which the vacuum energy is generally changed by an amount
typically many orders of magnitude larger than the observed
value~(\ref{Lambda:obs}). Many attempts have been made in the past
to try to understand the cosmological constant
problem~\cite{Weinberg:1988cp,Nobbenhuis:2006yf}. Most of these
attempts depend on new physical ideas, like new symmetries or
breaking the equivalence principle. However, ordinary quantum
field theory has a property that might shed some light on this
problem, namely the renormalization group (RG) .

\subsection{Renormalization Group}
\label{Renormalization Group}
 It is a well established fact~\cite{Peskin:1995ev}~\cite{Itzykson:1980rh} that upon
regularizing and renormalizing a divergence in quantum field
theory, one inevitably introduces a renormalization scale $\mu$.
Since this scale is arbitrary, physical observables should be
independent of this scale. This requirement is imposed by the
renormalization group equations. Solving these equations leads to
the RG improved theory, and the coupling constants of the theory,
including $\Lambda$, become a function of $\mu$. Since $\Lambda$
now runs with the energy scale $\mu$, one might hope that at
cosmological scales $\Lambda$ will run towards zero, independent
of its huge value in the ultraviolet where (\ref{QFT lambda}) is
calculated. Since the cosmological constant problem can be seen as
a conflict between scales: the ultraviolet given by $M_X$ where
$\Lambda_{\rho}$ is defined and the infrared where $\Lambda_{\rm
eff}$ is measured, it seems mandatory to study the RG behavior of
$\Lambda$.
\\
In~\cite{Shapiro:1999zt}~\cite{Shapiro:2000dz}~\cite{Shapiro:2006qx}
it is investigated how $\Lambda$ runs with $\mu$, when
$\Lambda_{\rho}$  is generated by the Standard Model particles.
$\Lambda_{\rm eff}$ as it is observed is then evaluated at
$\mu=H_0$, the Hubble parameter today. The result of these
calculations is that $\Lambda$ varies only very mildly (usually
logarithmically) with $\mu$ and the consensus is that this can
never take account of the missing 122 orders of magnitude, though
it
may make the finetuning issues somewhat less severe.\\
Another approach is to study the RG behavior of quantum
gravity~\cite{Reuter:1996cp,Reuter:2001ag,Reuter:2004nx}. While
this has produced very interesting results (discovery of an
ultraviolet fixed point), an explicit calculation of the effective
action is still lacking. The main reason for this lies of course
in the fact that we do not know what a renormalizable theory of
quantum gravity would look like. Another reason is that the
authors of~\cite{Reuter:1996cp,Reuter:2001ag,Reuter:2004nx} in the
course of the running project the effective action onto a certain
prescribed form (which typically contains $\Lambda$, $R$ and $R^2$
terms but nothing else). This form may be too restrictive since it
does not permit {\it e.g.} logarithmic dependence on spacetime
curvature, which plays an essential role in our one loop
investigation.
\\
 In this paper we consider in detail the RG behavior of
the cosmological constant in a manner that substantially differs
from the ideas mentioned above. We explicitly calculate the one loop
contributions to the effective action of an antisymmetric tensor
field with a mass given by the cosmological constant. Such a
field is motivated by nonsymmetric gravitational theories (see
section~\ref{Nonsymmetric gravitational theories}),
 but our results do not depend on its origins. For simplicity our
 calculations will be done in de Sitter space. Since de Sitter space
 is essentially empty space, with a cosmological constant, we consider our
 model exemplary for any space where the energy density is dominated by a cosmological
 constant.\\

 The Friedmann equation in de Sitter space is given by
\begin{equation}
    H^2 = \frac{\Lambda_c}{3} +
    \frac{\delta\Lambda_Q}{3}
\,, \label{Friedmann equation:Q}
\end{equation}
where $\delta\Lambda_Q$ stands for quantum loop contributions and
therefore depends on the renormalization scale $\mu$. An important
observation of this work is that loop corrections, apart from the
$\mu$ dependence, can dependent on the Hubble parameter (or more
generally on the curvature of space-time) in a rather complex
fashion, $\delta\Lambda_Q = \delta\Lambda_Q(\mu,H^2)$, such that
the self-consistent solution for $H^2$ in terms of $\Lambda_{\rm
c}$ as specified by Eq.~(\ref{Friedmann equation:Q}) is in general
of the form,
\begin{equation} \label{Fried: effective}
H^2 = \Lambda_{\rm eff}(\Lambda_{\rm c},\mu)/3,
\end{equation}
 where $\Lambda_{\rm eff}$ can be substantially different from $\Lambda_{\rm c}$
of the original theory.\\
An important question is at what scale $\mu$ one should evaluate
the quantum corrections. In other words: what is our ultraviolet
scale and what is the infrared scale where we measure
$\Lambda_{eff}$? It is often argued in the literature that a
natural infrared scale for $\mu$ is the Hubble scale. This means
that the Hubble length $H^{-1}$ is taken to be the longest
relevant wavelength. Causality arguments are sometimes invoked to
justify that. The ultraviolet scale is then given by $m_p$ or
$M_X$. We however do not agree with this approach. In accelerating
space times modes grow 'super-Hubble' and there is no reason to
assume that these modes do not have an effect on the geometry or
are unobservable. Therefore a proper treatment of the infrared
behavior of the theory must take these modes into account and one
should let $\mu$ run down to the energy of the largest
super-hubble mode, which is much less then $H$. As for the
ultraviolet scale the problem with choosing $m_p$ (or $M_X$) is
that, while $\mu$ runs to the infrared, the ratio $m_p/\mu$
becomes huge. This usually spoils the trustability of the one loop
expansion. Because of these reasons, we define our ultraviolet
theory at $\mu=H$, the Hubble parameter. This makes sense, since
not only is the scale $H$ naturally present in the theory, it is
also because of the Friedmann equation (\ref{Friedmann
equation:Q}) a natural UV scale. This is because the ultraviolet
$\Lambda_c$ is huge (see equation (\ref{QFT lambda})).
Furthermore, if we decrease $\mu$, we find that also $H$
decreases. Therefore the ratio $H/\mu$ stays closer to one then
for any other choice of
initial $\mu$, which makes the RG results more trustable.\\

As a last remark, we point out that a substantial difference
between $\Lambda_{\rm eff}$ and $\Lambda_c$ can only be obtained
if the quantum radiative effects are comparable to the tree level
cosmological constant. While this necessarily makes the loop
expansion problematic, we are not aware of any fundamental reason
why this compensation should not take place. An example of such a
behavior -- dubbed attractor mechanism -- is considered in
Ref.~\cite{Prokopec:2006yh}, where quantum loop matter
fluctuations generate a term which exhibits logarithmic dependence
on the Hubble parameter, and which compensates the tree level
cosmological term. The important difference between
Ref.~\cite{Prokopec:2006yh} and the work at hand is that the
compensation mechanism in~\cite{Prokopec:2006yh} is realised {\it
via} the dynamics of a scalar field in an effective potential
generated by quantum loop corrections, while here the compensation
occurs as a result of sending the RG scale
to zero.\\

\subsection{Nonsymmetric gravitational theories}
\label{Nonsymmetric gravitational theories}

 Nonsymmetric gravitational theories (NGTs) are extensions of
general relativity (GR), where the standard axiom that the metric
is symmetric is
dropped~\cite{Moffat:1994hv}~\cite{Clayton:1995yi}~\cite{Janssen:2006jx}.
One can therefore make a decomposition of the general metric
$c_{\mu\nu}$:
\begin{equation}
    c_{\mu\nu}=g_{\mu\nu}+ B_{\mu\nu}\qquad g_{\mu\nu}\equiv
    c_{(\mu\nu)}\qquad B_{\mu\nu}\equiv c_{[\mu\nu]},
\end{equation}
where $(\cdot)$ and $[\cdot]$ indicate normalized symmetrization
and anti-symmetrization, respectively. One of the reasons why such
a theory is interesting to study, is that it provides a natural,
geometric source for torsion~\cite{Mao:2006bb}. Another
interesting property is that the massive antisymmetric tensor
field might act as the dark matter component in the
universe~\cite{Prokopec:2006kr}~\cite{Prokopec:2005fb}~\cite{Valkenburg}.
Since GR is a highly successful it makes sense to consider NGT in
the limit of a small $B$-field. In ~\cite{Janssen:2006nn} it was
shown that the only Lagrangian that does not lead to an unstable
field evolution is given by (for the more precise form of this
Lagrangian see Eq.~(\ref{lagrangian}))
\begin{equation}\label{lagNGT}
    \mathcal{L}=\sqrt{-g}\bigg[Q(R-2\Lambda)+\frac{b}{144}R^2-\frac{1}{12}F^2-\frac{1}{4}m^2
    B^2\Bigg]+\mathcal{O}(B^3).
\end{equation}
Here the curvature term $R$ refer to the curvature of the GR
background and $F_{\mu\nu\rho}$ is the field strength associated
with $B_{\mu\nu}$. The mass term is given by
\begin{equation}
           m^2=\omega\Lambda+\frac{\chi}{12} R +\frac{\rho}{144 m_p^2} R^2,
\end{equation}
with $\omega$, $\chi$ and $\rho$ undetermined constants from the
theory. The cosmological constant, as it enters (\ref{lagNGT}), is
the sum of the geometric ($\Lambda_g$) and matter
($\Lambda_{\rho}$) contributions. The fact that the mass of the
$B$ field is proportional to the cosmological term is extremely
interesting, since in principle it induces a large back-reaction
on the total energy of the vacuum. This is the question we
investigate in detail in this paper.
\\
Most of the conclusions of this paper however are not dependent on
the fact that we are dealing with antisymmetric tensor fields. For
example a scalar field with a mass proportional to $\Lambda$, as might arise in
Kaluza-Klein type theories, would almost certainly lead to
similar conclusions.

\subsection{Overview}
The contents of our paper are as follows. In section
\ref{Propagators in de Sitter space}, we discuss some properties
of de Sitter space and show how to derive the propagator for the
$B$-field. In section \ref{section effective action} we derive the
one loop effective action and in section \ref{section
renormalization} we renormalize the theory and RG improve it. In
section \ref{sfried} we calculate the Friedmann equation and show
how the $B$-field loops alter the effective vacuum energy. We
discuss and summarise our results in section \ref{section discus}.

\section{Propagators in de Sitter space}
\label{Propagators in de Sitter space}

In this section we first review basic properties of de Sitter
space and then we sketch a standard derivation for the de Sitter
invariant scalar propagator. Having done this, we show how to
construct the propagator for a massive antisymmetric tensor field
in de Sitter space. We show that the propagator can be written as
the sum of two scalar propagators with different amounts of
conformal coupling. This is similar as in the case of the photon
propagator~\cite{Kahya:2005kj,Tsamis:2006gj} in de Sitter space.
In this section we raise and lower indices with $\eta^{\mu\nu}$
and $\eta_{\mu\nu}$, respectively, where
$\eta_{\mu\nu} = {\rm diag}(-1,1,1,..)$ is the Minkowski metric in
D space-time dimensions.

\subsection{de Sitter space}

 For pedagogical reasons we begin by introducing four dimensional de Sitter
space (later we shall work with general D dimensional de Sitter space),
which is the hypersurface given by the
equation~\cite{Spradlin:2001pw}~\cite{Prokopec:2006yh}
\begin{equation}
    -X_0^2+X_1^2+X_2^2+X_3^2+X_4^2=\frac{1}{H^2}
\end{equation}
embedded in 5-dimensional Minkowski space-time, where $H$ is
the Hubble parameter. The isometry group of de Sitter
space, $SO(1,4)$, is manifest in this embedding. We shall use
flat coordinates, which cover only half of the de Sitter
manifold, given by ($i=1,2,3$)
\begin{equation}
    \begin{split}
        X_0&=\frac{1}{H}\sinh(Ht)+\frac{H}{2}x_i x^i e^{H t},\\
        X_i&=e^{H t}x_i,\\
        X_4&=\frac{1}{H}\cosh(Ht)-\frac{H}{2}x_i x^i e^{H t},\\
        -\infty&<t,x_i<\infty.
    \end{split}
\label{coordinates}
\end{equation}
In these coordinates the metric reads
\begin{equation}
    g_{\mu\nu}=\text{diag}(-1,a^2,a^2,a^2)\,,\qquad\qquad a=e^{H t}
\,,
\end{equation}
which we can write in conformal form by changing coordinates to
conformal time $\eta$ defined as $ad\eta=dt$:
\begin{equation}
    g_{\mu\nu}=\text{diag}(-a^2,a^2,a^2,a^2)=a^2\eta_{\mu\nu},\qquad
    a=-\frac{1}{H\eta}
, \qquad\eta<0
\,.
 \label{conformal}
\end{equation}
We define the de Sitter invariant distance
functions~\cite{Allen:1985ux}~\cite{Allen:1987tz}
\begin{equation}
Z(X;X')=H^2\sum_{A,B=0}^4\eta_{AB}X^AX^{B\prime} =1-\frac{1}{2}Y(X;X')
\,.
\end{equation}
In conformal coordinates~(\ref{coordinates}) these functions read
\begin{equation} \label{dsinvar}
    \begin{split}
        z(x;x')&=1-\frac{1}{2}y(x;x')\\
        y(x;x')&=a a' H^2\Delta x^2(x;x')\\
        \Delta
        x^2(x;x')&=-(|\eta-\eta'|-i\epsilon)^2+||\vec{x}-\vec{x}'||^2,
    \end{split}
\end{equation}
where $Y(X;X')=y(x;x')$, $Z(X;X')=z(x;x')$, $a=a(\eta)$ and
$a'=a(\eta')$ are functions given in (\ref{conformal}) and
$\epsilon>0$ refers to the Feynman (time ordered) pole
prescription. The function $y=y(x;x')$ is related to the invariant
length $\ell=\ell(x;x')$ between points $x$ and $x'$ as,
\begin{equation}
 y(x;x') = 4\sin^2\left(\frac12 H\ell(x;x')\right)
\,.
\label{dS_inv_ell}
\end{equation}

\subsection{Scalar propagator in de Sitter space}

 The de Sitter invariant scalar propagator for a massive scalar field
is the expectation value
\begin{equation}
    i\Delta(x;x')=\langle x|\frac{i}{\sqrt{-g}(\Box-m^2-\xi R_D)}|x'\rangle
,
\label{prop1}
\end{equation}
where $R_D$ is the $D$-dimensional Ricci scalar, which in de
Sitter space-time is given by $R_D=D(D-1)H^2$. $\Box$ is the
d'Alembertian. The propagator~(\ref{prop1}) satisfies the following
Klein-Gordon equation
\begin{equation} \label{KG}
    \sqrt{-g}(\Box-m^2-\xi R_D) i\Delta(x;x')=i\delta^D(x-x')
,
\end{equation}
where $\delta^D$ is the $D$-dimensional Dirac delta. The de Sitter
invariant form of (\ref{KG}) is
\begin{equation} \label{KGdsinvar}
    \Big[(1-z^2)\frac{d^2}{dz^2}-D z\frac{d}{dz}-\frac{m^2+\xi
    R_D}{H^2}\Big]i G(y)=\frac{i\delta^D(x-x')}{a^D H^2},
\end{equation}
where the invariant propagator is defined as $i
G(y)=i\Delta(x;x')$ and we used
\begin{equation}
    \begin{split}
        \partial_\mu&=(\partial_\mu z)\frac{d}{dz}=-\frac{1}{2}H
        a\Big(\delta^0_\mu y+2 a' H\Delta x_\mu)\frac{d}{dz}\\
        \Box&=(\sqrt{-g})^{-1}\partial_\mu
        g^{\mu\nu}\sqrt{-g}\partial_\nu
           = \frac{1}{a^D}\partial_\mu a^{D-2}\eta^{\mu\nu}\partial_\nu
        \\
        &=H^2(1-z^2) \frac{d^2}{dz^2}
          + \Big(a^{-2}\eta^{\mu\nu}\partial_\mu\partial_\nu
        z-a^{-1}(D-2)H(\partial_0 z)\Big)\frac{d}{dz}
\,,
    \end{split}
\end{equation}
where $z$ and $y$ are defined in (\ref{dsinvar}).
Eq.~(\ref{KGdsinvar}) is a hypergeometric equation, whose
general solution is given in terms of hypergeometric functions
\begin{equation}
    iG(y)=c_1{\;}_2F_1\Big(
                \frac{D-1}{2}+\nu,\frac{D-1}{2}-\nu;\frac{D}{2};1-\frac{y}{4}
                     \Big)
        +c_2{\;}_2F_1\Big(
                  \frac{D-1}{2}+\nu,\frac{D-1}{2}-\nu;\frac{D}{2};\frac{y}{4}
                     \Big),
\label{scalar_prop_gen}
\end{equation}
where
\begin{equation}
    \nu=\Big(\frac{(D-1)^2}{4}-\frac{m^2+\xi
    R_D}{H^2}\Big)^{1/2}.
\end{equation}
The constants $c_1$ and $c_2$ are uniquely fixed if we require
that near the lightcone the solution reduces to the Hadamard form,
while there is no singularity at the antipodal lightcone (which
would lead to $\alpha$-vacua~\footnote{$\alpha$-vacua are probably unphysical
since the propagator equation requires an additional $\delta$-function source
at the antipodal
point~\cite{Allen:1985ux}~\cite{Spradlin:2001pw}~\cite{Mottola:1984ar},
henceforth we shall not consider them in this paper.}).
This means we require:
\begin{equation}
    \begin{split}
    \lim_{y\rightarrow
    0}iG(y)&=\frac{H^{D-2}}{(4\pi)^{D/2}}\Gamma\Big(\frac{D}{2}-1\Big)
    y^{1-\frac{D}{2}}\\
    \lim_{y\rightarrow
    4}iG(y)&=0,
    \end{split}
\end{equation}
which uniquely specifies both constants in
Eq.~(\ref{scalar_prop_gen}) and we arrive at the well known
Chernikov-Tagirov scalar propagator in de Sitter
space~\cite{Chernikov:1968zm}
\begin{equation}
iG(y)=\frac{H^{D-2}}{(4\pi)^{D/2}}\frac{\Gamma\Big(\frac{D-1}{2}+\nu\Big)
       \Gamma\Big(\frac{D-1}{2}-\nu\Big)}{\Gamma\Big(\frac{D}{2}\Big)}
        {\;}_2F_1\Big(\frac{D-1}{2}+\nu,\frac{D-1}{2}-\nu;
                    \frac{D}{2};1-\frac{y}{4}
               \Big)
\,.
\label{scalarprop}
\end{equation}

\subsection{Calculating the $B$-field propagator}

 In this section we calculate the propagator for the
anti-symmetric tensor field ($B$-field) in de Sitter space. Our
starting lagrangian is~\cite{Janssen:2006nn}
\begin{equation} \label{lagrangian}
    \mathcal{L}=\sqrt{-g}\Bigg(Q(R-2\Lambda)+\frac{b}{144}R^2
    -\frac{1}{12}g^{\alpha\mu}g^{\beta\nu}g^{\gamma\rho}
                       F_{\alpha\beta\gamma}F_{\mu\nu\rho}
     -\frac{1}{4}m^2g^{\alpha\mu}g^{\beta\nu}B_{\alpha\beta}B_{\mu\nu}\Bigg)
\,,
\end{equation}
where $R$ is the Ricci scalar, $g = {\rm det}[g_{\mu\nu}]$ and
\begin{equation}
\begin{split}
    m^2&= \omega\Lambda+\chi H^2
\,,\qquad
    Q = \frac{1}{16\pi G}
\\
    F_{\alpha\beta\gamma}&=\partial_\alpha
    B_{\beta\gamma}+\partial_\beta
    B_{\gamma\alpha}+\partial_\gamma B_{\alpha\beta}
\,.
    \end{split}
\label{m2_Fabc}
\end{equation}
Notice that in principle there also is a $H^4$ contribution to
$m^2$ however there is nothing that forbids us to set this
contribution to zero and we do so for simplicity. We will focus on
the B-field contribution (remember that we raise/lower indices
with $\eta^{\mu\nu}/\eta_{\mu\nu}$ and work in the conformal
coordinates of de Sitter space (\ref{conformal}))
\begin{equation}
\begin{split}
        \mathcal{L}_B&=a^D\Big(-\frac{1}{12}a^{-6}F_{\alpha\beta\gamma}F_{\mu\nu\rho}\eta^{\alpha\mu}\eta^{\beta\nu}\eta^{\gamma\rho}-\frac{1}{4}a^{-4}m^2B_{\alpha\beta}B_{\mu\nu}\eta^{\alpha\mu}\eta^{\beta \nu}\Big)\\
        &=\frac{1}{4}B_{\nu\rho}\partial_\mu\Big(a^{D-6}\Big[\partial_\alpha
        B_{\beta\gamma}+\partial_\beta
        B_{\gamma\alpha}+\partial_\gamma
        B_{\alpha\beta}\Big]\Big)\eta^{\alpha\mu}\eta^{\beta\nu}\eta^{\gamma\rho}-\frac{1}{4}a^{D-4}m^2B_{\alpha\beta}B_{\mu\nu}\eta^{\alpha\mu}\eta^{\beta\nu}\Big)\\
        &=\frac{1}{4}B^{\beta\gamma}\partial^\mu(a^{D-6}\partial_\mu
        B_{\beta\gamma})+\frac{1}{2}a^{D-6}\Big(\partial^\alpha
        B_{\gamma\alpha}-(D-6)a H B_{\gamma
        0}\Big)\Big(\partial_\beta B^{\gamma\beta}-(D-6)a H B^{0
        \gamma}\Big)\\
        &\qquad +\frac{1}{2}(D-6)H^2 a^{D-4}B_{\gamma
        0}B^{0\gamma}-\frac{1}{4}a^{D-4}m^2B_{\alpha\beta}B_{\mu\nu}\eta^{\alpha\mu}\eta^{\beta\nu},
    \end{split}
\end{equation}
where in the second line we have dropped a total derivative. Next
we add a gauge fixing term~\footnote{The terminology 'gauge fixing'
is a bit misleading, since the theory has no explicit gauge
invariance. However, one still is able to add a 'gauge fixing'
term. It plays the r$\hat{\text{o}}$le of a Lagrange multiplier,
imposing the Lorentz condition. See
e.g~\cite{Embacher:1986pt}~\cite{Ruegg:2003ps}.}:
\begin{equation}    \mathcal{L}_{GF}=-\frac{1}{2}a^{D-6}\Big(\partial^\alpha
    B_{\gamma\alpha}-(D-6)a H B_{\gamma 0}\Big)\Big(\partial_\beta
    B^{\gamma\beta}-(D-6)a H B^{0\gamma}\Big)
\end{equation}
to obtain
\begin{equation}
\begin{split}
    \mathcal{L}_{B-GF}&=\frac{1}{4}\Big(B^{\beta\gamma}\partial^\alpha(a^{D-6}\partial_\alpha)B_{\beta\gamma}-a^{D-4}m^2B^{\beta\gamma}B_{\beta\gamma}-2(D-6)H^2a^{D-4}B_{\gamma
        0}B^{\gamma 0}\Big)\\
      &=-\frac{1}{2}B^{\mu\nu}\mathcal{D}^{\rho\sigma}_{\mu\nu}B_{\rho\sigma}
\,,
    \end{split}
\end{equation}
where we have defined
\begin{equation} \label{kineticD}
    \mathcal{D}^{\rho\sigma}_{\mu\nu}=-\frac{1}{2}\Bigg[(\partial^\alpha
    a^{D-6}\partial_\alpha-a^{D-4}m^2)\delta^{[\rho}_\mu\delta^{\sigma]}_\nu-2(D-6)H^2
    a^{D-4}\delta^0_{[\nu}\bar{\delta}_{\mu]}^{[\rho}\delta^{\sigma]}_0\Bigg]
\end{equation}
and
\begin{equation}
    \begin{split}
        \bar{\delta^\mu_\nu}&=\delta^\mu_\nu-\delta^\mu_0\delta^0_\nu
\,,\qquad
        \bar\eta_{\mu\nu}=\eta_{\mu\alpha}\bar\delta^\alpha_\nu
                          = \eta_{\mu\nu}+\delta^0_\mu\delta^0_\nu
\,.
        \end{split}
\end{equation}
 The propagator is given by
\begin{equation}
 i\mathcal{D}^{\rho\sigma}_{\mu\nu}[{}_{\rho\sigma}\Delta_{\alpha\beta}](x;x')
       =i\eta_{\alpha[\mu}\eta_{\nu]\beta}\delta^D(x-x')
\,.
\end{equation}
We use the following ansatz for the propagator (the subscripts 2
and 3 will become clear shortly)
\begin{equation}
{}_{\rho\sigma}\Delta_{\alpha\beta}(x;x')=-2a^2a'^2\Big[\bar{\eta}_{\alpha[\rho}\bar{\eta}_{\sigma]\beta}\Delta_2(x;x')+\delta^0_{[\sigma}\bar{\eta}_{\rho][\alpha}\delta^0_{\beta}]\Delta_3(x;x')\Big].
\end{equation}
 With the help of the identity
\begin{equation}
 \begin{split}       (\partial^\alpha a^{D-6}\partial_\alpha)(2a^2
        a'^2\Delta)&=a^D(\partial^\sigma
        a^{D-2}\partial_\sigma-2(D-3)H^2)\Delta\\
        &=a^D(\Box-2(D-3)H^2)\Delta
    \end{split}
\end{equation}
we find that
\begin{equation}
\begin{split}
    \Big(\Box-2(D-3)H^2-m^2\Big)i\Delta_2(x;x')&=a^{-D}i\delta^D(x-x')\\
    \Big(\Box-3(D-4)H^2-m^2\Big)i\Delta_3(x;x')&=a^{-D}i\delta^D(x-x').
    \end{split}
\end{equation}
So we see that, similarly to the case of the photon propagator
in~\cite{Kahya:2005kj}, that we can write the $B$-field propagator
in terms of scalar propagators with various types of coupling to
the Ricci scalar.

Our $\Delta_2$ corresponds to $\Delta_C$
of~\cite{Kahya:2005kj}~\cite{Tsamis:2005je}. In fact the
propagators $A$, $B$ and $C$ of these references correspond to
$n=0,1,2$ respectively of
\begin{equation}
    \Big(\Box-n(D-n-1)H^2-m^2\Big)i\Delta_n(x;x')=a^{-D}i\delta^D(x-x').
\end{equation}
This equation is solved by (\ref{scalarprop}), with
\begin{equation}
    \begin{split}
    \nu&\rightarrow
    \nu_n=\Big(\frac{(D-2n-1)^2}{4}-\frac{m^2}{H^2}\Big)^{1/2}\\
    iG(y)&\rightarrow iG_n(y) \,. \label{nu}
    \end{split}
\end{equation}

\section{The Effective Action}\label{section effective action}

Since the $B$-field appears quadratically in our lagrangian
(\ref{lagrangian}), it can be integrated out. Thus we get (up to
an irrelevant normalization constant)
\begin{equation}
\exp[i\Gamma(g_{\mu\nu})]=\int \mathcal{D} B \exp[i
S(g_{\mu\nu},B_{\mu\nu})]= \exp[i
    S_{HE}(g_{\mu\nu})]\frac{1}{\sqrt{Det[D^{\mu\nu\rho\sigma}]}}
\end{equation}
and therefore
\begin{equation}
\Gamma(g_{\mu\nu})=S_{HE}(g_{\mu\nu})+\frac{i}{2}
    \text{Tr}\ln(\mathcal{D}^{\mu\nu\rho\sigma})+\text{higher
    loops},
\end{equation}
where the trace refers to the spacetime integration $\int d^D
x\equiv\int_x$ and the contraction over the Lorentz indices. Using
(\ref{kineticD}) we find
\begin{equation}
\begin{split}
\Gamma_B&\equiv\frac{i}{2}\text{Tr}\ln(D^{\mu\nu\rho\sigma})\\
&=\frac{i}{2}\int_x
\Big(\eta_{\rho[\mu}\eta_{\nu]\sigma}\ln\Big[-\frac{1}{2}\Big((\partial^\alpha
    a^{D-6}\partial_\alpha-a^{D-4}m^2)\eta^{\mu[\rho}\eta^{\sigma]\nu}
-2(D-6)H^2a^{D-4}\delta_0^{[\nu}\bar{\eta}^{\mu][\rho}\delta^{\sigma]}_0\Big)\Big]\Big)\,.
    \end{split}
\end{equation}
The standard
technique~\cite{Itzykson:1980rh}~\cite{Birrell:1982ix} to get rid
of the log is to take a derivative with respect to the mass and
then write $\Gamma$ in terms of the propagator. In our case
however, we have to take a log of two expressions which are,
because of the tensorial structure, orthogonal. To take this
properly into account we need to take both the derivative with
respect to the mass and with respect to $H^2$. We obtain
\begin{equation} \label{gammab}
   \begin{split}
\Gamma_B=&\frac{i}{2}\int_x\Bigg[\int dm^2
\Big(\eta_{\rho[\mu}\eta_{\nu]\sigma}[{}^{\rho\sigma}\Delta^{\alpha\beta}](x;x)(\frac{1}{2}a^{D-4})\delta^{[\mu}_\alpha\delta^{\nu]}_\beta\Big)\\
    &\quad-\int
    dH^2\Big(\eta_{\rho[\mu}\eta_{\nu]\sigma}[{}^{\rho\sigma}\Delta^{\alpha\beta}](x;x)((D-6)a^{D-4}\delta^{[\nu}_0\bar{\eta^{\mu]}_{[\alpha}}\delta^0_{\beta]})\Big)\Bigg]\\
    &=-\frac{1}{4}\int_x\Bigg[\int dm^2
    a^D\Big((D-1)(D-2)i\Delta_2(x;x)-(D-1)i\Delta_3(x;x)\Big)\\
    &\qquad\qquad+\int dH^2 a^D(D-1)(D-6)i\Delta_3(x;x)\Bigg]\\
    &=-\frac{1}{4}\int_x a^D\Bigg[\int
    dw(D-1)H^2\Bigg((D-2)i\Delta_2(x;x)-i\Delta_3(x;x)\Bigg)\\
    &\qquad+\int dH^2(D-1)\Bigg((D-2)w
    i\Delta_2(x;x)+((D-6)-w)i\Delta_3(x;x)\Bigg)\Bigg]\\
    &\equiv \Gamma_w+\Gamma_H,
    \end{split}
\end{equation}
where
\begin{equation}    w\equiv\frac{m^2}{H^2}.
\end{equation}
Since at the coincidence limit only the $y^0$ term of the
propagators contribute (the other powers are $D$-dependent powers
and do not contribute in dimensional regularization), we use the
following expression for the scalar propagators.
\begin{equation}\label{coincidence prop}
    i\Delta_n(x;x)\Big|_{y^0-\text{term}}
    =\frac{H^{D-2}}{(4\pi)^{D/2}}
      \Gamma\Big(1-\frac{D}{2}\Big)
      \frac{\Gamma\Big(\frac{D-1}{2}+\nu_n\Big)
             \Gamma\Big(\frac{D-1}{2}-\nu_n\Big)}
           {\Gamma\Big(\frac{1}{2}+\nu_n\Big)\Gamma
                \Big(\frac{1}{2}-\nu_n\Big)},
\end{equation}
with $\nu_n$ defined in (\ref{nu}). In order to renormalize the
theory, we need to split the effective action (\ref{gammab}), with
propagators given by (\ref{coincidence prop}), in its finite and
infinite parts. This separation is done in Appendix~\ref{appendix
effective action}. The final result is
\begin{equation}\label{deltaL}
   \begin{split}
        \Gamma_B=&-\int_x\sqrt{-g}\Bigg[\\
        &\frac{D-1}{4(4\pi)^{D/2}}\Gamma(1-D/2)
     \Bigg(H^4\frac{1}{2D}
          \Big[8(D-5)(D-6)+2(24+D(D^2-6D+4))\chi+(D-3)(D+4)\chi^2
          \Big]
      \\
        &\qquad+H^2\omega\Lambda \Big(4 + D^2+D(\chi-6) -
        \frac{12(\chi-2)}{D}+\chi\Big)+(\omega\Lambda)^2 \frac{(D-3)(D+4)}{2D}\Bigg)\mu^{D-4}\\
       &+\frac{3}{64\pi^2}\Bigg(\Big((\omega\Lambda)^2 +2(1+\chi)\omega\Lambda H^2+2(1+\chi+\frac{\chi^2}{2})H^4\Big)\ln\Big(\frac{\omega\Lambda +\chi H^2}{\mu^2}\Big)-\frac{31}{15}H^4\ln\Big(\frac{\omega\Lambda +\chi H^2}{H^2}\Big)\\
       &\qquad-\frac{(\omega\Lambda)^2 }{4}+\frac{H^2\omega\Lambda }{2}(1-\chi)+\frac{\chi
       H^4}{2}(1-\frac{\chi}{2})+\frac{13}{10}H^4\Bigg)\Bigg]+\mathcal{O}\Big(H^4\frac{H^2}{m^2}\Big)\\
       \equiv&\int_x\delta\mathcal{L}.
   \end{split}
\end{equation}
In this calculation we used the approximation that
\begin{equation}\label{assumption}
    \frac{H^2}{\omega\Lambda+\chi H^2}\ll 1,
\end{equation}
which we justify in section \ref{sassumpproof}.
\\
 The effective action(\ref{deltaL}) can be simplified further
in the special case when
$\chi\leq\mathcal{O}(1)$, since then our approximation
(\ref{assumption}) is equivalent to ${H^2}/(\omega\Lambda)\ll
1$. In this case we should replace the logs in (\ref{deltaL}):
\begin{equation} \label{logreplacements}
    \begin{split}
        \ln\Big(\frac{\omega\Lambda +\chi H^2}{H^2}\Big)&\rightarrow
        \ln\Big(\frac{\omega\Lambda}{H^2}\Big)+\frac{\chi H^2}{\omega\Lambda}-\frac{1}{2}\Big(\frac{\chi H^2}{\omega\Lambda}\Big)^2\\
        \ln\Big(\frac{\omega\Lambda +\chi H^2}{\mu^2}\Big)&\rightarrow
        \ln\Big(\frac{\omega\Lambda}{\mu^2 }\Big)+\frac{\chi H^2}{\omega\Lambda}-\frac{1}{2}\Big(\frac{\chi H^2}{\omega\Lambda}\Big)^2
\,.
    \end{split}
\end{equation}

\section{Renormalization}\label{section renormalization}

 The Lagrangian we wish to renormalize is (remember that in de
 Sitter space $R=12 H^2$)
\begin{equation}
    \mathcal{L}=\sqrt{-g}\Big(Q(12 H^2-2\Lambda)+b
    H^4+\delta\mathcal{L}\Big)
\end{equation}
and to do so we add the following counterterms
\begin{equation}
    \mathcal{L}_{CT}=\sqrt{-g}\Big(\Lambda_c+Q_c H^2+b_cH^4\Big).
\end{equation}
We use a regularization scheme in which the counterterms
remove all non-log terms from $\delta\mathcal{L}$. This scheme has
the advantage that the equations of motion become relatively
simple. The results in other regularization schemes are related to
this regularization scheme by a finite shift in the coupling constants.
A potential disadvantage of this scheme is that it is not
immediately clear what are the physical Newton and
cosmological constant. This issue we shall study in more
detail in section \ref{sparameters}. Using our regularization
scheme the lagrangian becomes
\begin{equation} \label{Lren}
\begin{split}
    \mathcal{L}_{\rm ren}=\sqrt{-g}\Bigg(&Q(12 H^2-2\Lambda)+bH^4
\\
 &-\frac{3}{64\pi^2}\Bigg[\Big((\omega\Lambda)^2 +2(1+\chi)\omega\Lambda H^2
                         +2(1+\chi+\chi^2/2)H^4\Big)
                            \ln\Big(\frac{\omega\Lambda +\chi H^2}{\mu^2}\Big)
\\
    &-\frac{31}{15}H^4\ln\Big(\frac{\omega\Lambda +\chi
    H^2}{H^2}\Big)\Bigg]+\mathcal{O}\Big(\frac{H^6}{m^2}\Big)\Bigg)\,.
\end{split}
\end{equation}
Notice that if $\chi\sim\mathcal{O}(1)$, Eq.~(\ref{assumption})
implies that the replacement~(\ref{logreplacements}) is
applicable, one should only replace the logs, since the other
terms generated by (\ref{logreplacements}) are removed in our
regularization scheme.

\subsection{RG improvement}
\label{RG improvement}

 The renormalized lagrangian~(\ref{Lren}) still contains
the arbitrary mass-scale $\mu$. The dependence on this  scale is
unphysical, since it is introduced by the counterterms. Removing
the $\mu$ dependence is tantamount to the renormalization group
improvement~\cite{Coleman:1973jx}~\cite{Kleinert:2001ax}~\cite{Itzykson:1980rh}~\cite{Peskin:1995ev}.
From
\begin{equation}
    \mu\partial_\mu\mathcal{L}_{\rm ren}
  =\sqrt{-g}\frac{3}{32\pi^2}
   \left(
         (\omega\Lambda)^2 + 2(1+\chi)\omega\Lambda H^2
       + 2\Big(1+\chi+\frac12\chi^2\Big)H^4
   \right)
\label{mudmuL}
\end{equation}
we read off the beta-functions
\begin{equation}
\begin{split}
 &\beta_\Lambda = \frac{3}{64\pi^2}
                  \frac{(\omega\Lambda)^2 }{Q}
                + \frac{1+\chi}{64\pi^2}\frac{\omega\Lambda^2 }{Q}
\\
    &\beta_Q = -\frac{1+\chi}{64\pi^2}\omega\Lambda
\\
    &\beta_b = -\frac{3}{16\pi^2}\Big(1+\chi+\frac12\chi^2\Big)
\,.
\end{split}
\label{betas}
\end{equation}
 The idea is now to 'improve' our lagrangian by imposing the
renormalization group equation, which in our case (where wave
function renormalization can be neglected) can be well
approximated by the Callan-Symanzik equation
\begin{equation}
   (\mu\partial_\mu+\beta_\Lambda\partial_\Lambda
      +\beta_Q\partial_Q+\beta_b\partial_b)\mathcal{L}_{\rm ren}
    =0.
\label{CS equation}
\end{equation}
We solve this equation by the method of
characteristics~\cite{Kleinert:2001ax}~\cite{Faux} (see also
Appendix~\ref{ap method of char}), which means that we need to make the
substitution
\begin{equation}\label{replacement}
    \mathcal{L}_{\rm ren}(\Lambda,Q,b,\mu)\rightarrow
    \mathcal{L}_{\rm ren}(\Lambda[t],Q[t],b[t],\mu[t]),
\end{equation}
where $\Lambda[t]$, $Q[t]$ and $b[t]$ are the solutions of the
differential equations
\begin{equation} \label{diffeq}
    \frac{d}{dt}\Lambda[t]=\beta_\Lambda
\,,\qquad
   \frac{d}{dt}Q[t]=\beta_Q
\,,\qquad
    \frac{d}{dt}b[t]=\beta_b
\,,
\end{equation}
and
\begin{equation}
    \mu[t]=\mu e^t.
\label{mu-t}
\end{equation}
$t$ is a parameter, independent of the couplings $\Lambda$, $Q$ or
$b$, that we choose to be
\begin{equation}\label{t}
    t=\ln\Big(\frac{H}{\mu}\Big)
\end{equation}
such that
\begin{equation}
    \mu[t]=H
\end{equation}
 This choice of $t$ is however not unique, and different
choices lead in principle to different RG improved effective
actions with different boundary conditions. However, choosing a
different $t$ differs from the effective
action~(\ref{RGimprovedlag}) only at higher order in the coupling
constants, and hence we believe that the results presented in this
work are generic. On the other hand, since in this work tree level
and one loop contributions are comparable, this question does
deserve further study.

Our motivation for the choice  (\ref{t}) is the following: first
of all the scale $H$ is naturally present in the theory. Moreover,
since $\Lambda$ in the ultraviolet is huge (see (\ref{QFT
lambda})), $t=0$ ($\mu=H$) gives a natural ultraviolet scale. One
could argue that $\mu=M_X$ or $\mu=m_p$ are also natural
ultraviolet scales. While this in principle is true, such a choice
presents the problem that when $\mu$ runs to zero, $t$ blows up.
This is problematic, since this means that the running parameters
hit the Landau pole (see (\ref{betasol})). Our choice (\ref{t}) on
the other hand does not present this problem, since we will show
that $H$ decreases when $\mu$ decreases in such a way that $t$
never blows up. Thus the choice (\ref{t}) has the advantage that
it results in trustable running from the far ultraviolet to the
far infrared.
\\
The solutions to the renormalization group equations (\ref{diffeq}
are given by
\begin{equation}\label{betasol}
    \begin{split}
  Q[t]
     &=Q_0\Big(1-2(2(1+\chi)+3\omega)\tilde Q_0^{-1}
                      \omega\Lambda_0t\Big)^{\frac{1+\chi}{2(1+\chi)+3\omega}}
\\
 \Lambda[t]
     &=\Lambda_0\Big(1-2(2(1+\chi)+3\omega)\tilde Q_0^{-1}\omega\Lambda_0 t
                 \Big)^{-\frac{1+\chi+3\omega}{2(1+\chi)+3\omega}}
\\
        b[t]&=b_0-\frac{3}{16\pi^2}\Big(1+\chi+\frac12\chi^2\Big)t,
    \end{split}
\end{equation}
where a subscript zero means evaluation of the parameter at $t=0$, and
\begin{equation}
\tilde{Q}_0=128\pi^2Q_0 = 8\pi m_p^2,
\end{equation}
where $m_p = 1.2\times 10^{19}~{\rm GeV}$ is the Planck scale.

 With the replacement (\ref{replacement}) we get the final RG
improved lagrangian
\begin{eqnarray}\label{RGimprovedlag}
 \mathcal{L}_{RG}
   &=& \sqrt{-g}\Bigg[Q[t](R-2\Lambda[t])+b[t]H^4
\\
   &-&\!\! \frac{3}{64\pi^2}\bigg((\omega\Lambda[t])^2
         + 2(1\!+\!\chi)\omega\Lambda[t]H^2
         + 2\Big(\!-\!\frac{1}{30}\!+\!\chi\!+\!\frac12\chi^2\Big)H^4\bigg)
             \ln\Big(\frac{\omega\Lambda[t]\!+\!\chi H^2}{H^2}\Big)
                        \Bigg]
    +\mathcal{O}\Big(\frac{H^6}{m^2}\Big).
\nonumber
\end{eqnarray}
This effective lagrangian is finite and independent on the scale
$\mu$ (in the sense of the Callan-Symanzik equation), and we use
it in the following to study how the one-loop $B$ field quantum
corrections influence the Friedmann equation in de Sitter space.

\subsection{Physical parameters and boundary conditions}
\label{sparameters}

Since $H$ appears in the logarithm of (\ref{RGimprovedlag}), it is
a priori not clear what we mean with our physical parameters.
Since we -- naturally -- do not want our model to spoil {\it e.g.}
solar system measurements, it is important to identify physical
parameters. We define the physical constants (denoted by subscript
$p$) in a standard way
\begin{equation} \label{paramaterdef}
    \begin{split}
   \Lambda[t]_p Q[t]_p &\equiv - \frac12\lim_{H\rightarrow 0}
                              \frac{\mathcal{L}}{\sqrt{-g}}
\\
   Q[t]_p& \equiv \frac{1}{12}
               \lim_{H\rightarrow 0}\frac{\partial}{\partial H^2}
                         \frac{\mathcal{L}}{\sqrt{-g}}
\\
  b[t]_p&\equiv
     \frac12\lim_{H\rightarrow 0}\Big(\frac{\partial}{\partial H^2}\Big)^2
                     \frac{\mathcal{L}}{\sqrt{-g}}
\,.
    \end{split}
\end{equation}
Our analysis in section \ref{sfried} shows that $\Lambda_p[t]$ is
not the relevant parameter that determines the expansion rate of
the universe. Instead, the relevant quantity which determines the
expansion rate is given by $\Lambda_{\rm eff}$, which is the self
consistent solution to the quantum Friedmann equation.

 In the following we assume $\omega\gg\chi+1$ and
$\chi\leq \mathcal{O}(1)$, such that we can apply (\ref{logreplacements})
(see also the note after equation (\ref{Lren})). In this limit,
the RG improved parameters (\ref{betasol}) become
\begin{equation}\label{betax-1}
    \begin{split}
  Q[t]&\simeq Q_0
\\
   \Lambda[t]&\simeq\frac{\Lambda_0}{1-6\omega^2(\tilde{Q}_0)^{-1}\Lambda_0 t}
\\
    b[t]&=b_0-\frac{3}{16\pi^2}\Big(1+\chi+\frac12\chi^2\Big)t\,.
    \end{split}
\end{equation}
 With these parameters, we calculate (\ref{paramaterdef}) for the
lagrangian (\ref{RGimprovedlag}) and obtain
\begin{equation} \label{Lambdap}
    \begin{split}
  Q_p[t]\Lambda_p[t]&=\lim_{H\rightarrow 0}\Big[Q[t]\Lambda[t]
         +\frac{3}{128\pi^2}\omega^2\Lambda[t]^2
                      \ln\Big(\frac{\omega\Lambda[t]}{H^2}\Big)\Big]
\\
        &= 2Q_0\Lambda[t] + \mathcal{O}\Big(\frac{1}{t^2},H^2\Big)
\,.
    \end{split}
\end{equation}
 To calculate the first derivative of ${\cal L}_{\rm RG}/\sqrt{-g}$,
first note that the derivative of the parameters $Q[t]$, $\Lambda[t]$ and
$b[t]$ can be easily obtained by noting, $\partial_t = -\mu\partial_\mu$,
which is evaluated in Eq.~(\ref{mudmuL}), whereby the potentially
 divergent $1/H^2$ terms cancel out. The result is,
\begin{equation} \label{Qp}
    \begin{split}
        Q_p[t]& = Q[t]
                - \frac{\omega\Lambda[t]}{128\pi^2}
                 \left[(1+\chi)\ln\left(\frac{\omega\Lambda[t]}{H^2}\right)
                       + \frac{\chi}{2}
                 \right]
                + \mathcal{O}\Big(\frac{1}{t},H^2\Big)\\
  & = Q_0\left[1-\frac{1+\chi}{3\omega}\right]
              + \mathcal{O}\Big(\frac{1}{t},H^2\Big)
\,,
    \end{split}
\end{equation}
where we used
\begin{equation}
    \ln\Big(\frac{\omega\Lambda[t]}{H^2}\Big)
         = \ln\Big(\frac{\omega\Lambda[t]}{\mu^2}\Big) -2t \sim - 2t
\,.
\end{equation}
 Equation (\ref{Qp})
implies that $Q_p$ is equal to $Q_0$ plus a small,
$t$-independent, correction (remember that we assumed
$\omega\gg\chi+1$). Since this means that the Newton constant does
not run with the scale $\mu$, standard gravitational tests are not
affected in this limit.
\\
Furthermore we find that, since $Q_p\simeq Q_0$,
Eq.~(\ref{Lambdap}) implies that
\begin{equation}
    \Lambda_p[t]\simeq 2\Lambda[t]
\,,
\end{equation}
and therefore it makes sense to put (see Eq.~(\ref{QFT lambda}))
\begin{equation}
    \Lambda_0 = M_X{^4}/m_p^2.
\end{equation}
 Unfortunately it does not seem to be
possible to make a similar statement for $b$. However this is of
no great concern to us, since the term $b_0 H^4$ does not
contribute to the Friedmann equation and thus it is not expected
to influence gravitational tests.

\section{The Friedmann equation}\label{sfried}

From (\ref{betasol}) it is clear that $\Lambda$ runs only
logarithmically with $\mu$ and this is never sufficient to get
$\Lambda[\mu]\rightarrow 0$ in a satisfying
manner~\cite{Nobbenhuis:2006yf}. The physically relevant quantity
however is the effective cosmological constant given by the
(modified) Friedmann equation
\begin{equation}
    H^2=\frac{\Lambda_{\rm eff}}{3}
\,,
\end{equation}
where $\Lambda_{\rm eff}$ may be very different from $\Lambda$.
To get the Friedmann equation we need to calculate
\begin{equation}
 \frac{\partial\mathcal{L}_{\rm RG}}{\partial g^{\mu\nu}}=0\rightarrow
-\frac{1}{2}\mathcal{L}_{\rm RG}/\sqrt{-g}+\frac{1}{4}H^2\frac{\partial
(\mathcal{L}_{\rm RG}/\sqrt{-g})}{\partial H^2}+\frac{1}{8}\frac{\partial
(\mathcal{L}_{\rm RG}/\sqrt{-g})}{\partial t}=0
\end{equation}
for the RG improved lagrangian (\ref{RGimprovedlag}). After some
algebra and dividing the result by $-3Q$ we obtain
\begin{equation} \label{fried}
    \begin{split}
       &H^2\Bigg\{1+(1+\chi)\frac{\omega\Lambda[t]}{\tilde{Q}[t]}
         -\bigg(1-(1+\chi+3\omega)
                \frac{\omega\Lambda[t]}{\tilde{Q}[t]}
          \bigg)
                \frac{\omega\Lambda[t]}{\tilde{Q}[t]}
              \bigg[(1+\chi)\ln\Big(\frac{\omega\Lambda[t]+\chi H^2}{H^2}\Big)
                   +1+\frac{\chi}{2}
              \bigg]
           \Bigg\}
\\
  &-\frac{\Lambda[t]}{3}
     \Bigg\{1-\frac{3}{2}\frac{\omega^2\Lambda[t]}{\tilde{Q}[t]}
          + \bigg(1-(1+\chi+3\omega)\frac{\omega\Lambda[t]}{\tilde{Q}[t]}\bigg)
                \frac{3\omega^2\Lambda[t]}{\tilde{Q}[t]}
          \bigg[\ln\Big(\frac{\omega\Lambda[t]+\chi H^2}{H^2}\Big) +\frac{1}{2}
          \bigg]
     \Bigg\}\\
        &+\frac{H^4}{\tilde{Q}[t]}
            \Bigg\{1+\chi+\frac12\chi^2
             +\frac{1}{3}
                 \bigg(
                      1-(1+\chi+3\omega)\frac{\omega^2\Lambda[t]}{\tilde{Q}[t]}
                 \bigg)
            \Bigg\}=0\,,
    \end{split}
\end{equation}
where we have removed a $H^4/m^2$ term, in agreement with our
approximation (\ref{assumption}). In order to find the effective
cosmological constant, we need to find the self consistent
solution of (\ref{fried}) for $H^2$.
\\

\subsection{Case 1: $\omega^2\Lambda[t]/\tilde{Q}[t]\ll 1$}
At first instance, this case might appear to be the most
interesting to look at. In this limit the Friedmann equation reads
 \begin{equation}
    H^2-\frac{\Lambda[t]}{3}+\alpha\frac{H^4}{\tilde Q[t]}=0\qquad\qquad
    \alpha\equiv \frac{31}{30}+\chi+\frac12\chi^2
,
 \end{equation}
which is solved by
\begin{equation} \label{case1fried}
    H^2=\tilde Q[t]
         \frac{\sqrt{1+\alpha\frac{4\Lambda[t]}{3\tilde Q[t]}}-1}{2\alpha}.
\end{equation}
Clearly there is no way in which (\ref{case1fried}) would lead to
$\Lambda_{\rm eff}\rightarrow 0$. To compare (\ref{case1fried})
with the standard form of the Friedmann equation, we assume
$\alpha\Lambda[t]\ll \tilde Q[t]$ and get
\begin{equation}
    H^2=\frac{\Lambda[t]}{3}
         \bigg[1-\frac{\alpha\Lambda[t]}{3\tilde Q[t]}
            + \mathcal{O}\Big(\frac{\alpha\Lambda[t]}{3\tilde Q[t]}\Big)^2\,
         \bigg].
\end{equation}
Therefore we see that we get a small correction to the Friedmann
equation.\\
On the other hand, one could assume that
$\alpha\Lambda[t]\gg\tilde Q[t]$. In this case
Eq.~(\ref{case1fried})implies that $H^2\simeq \sqrt{\tilde
Q[t]\Lambda[t]/(3\alpha)}$, such that $H^2$ is given by the
geometric mean between $\Lambda[t]$ and the Planck scale $\tilde
Q_0 = 8 \pi m_p^2$. This clearly does not help us in addressing
the cosmological constant problem. Moreover, since in this limit
perturbative calculations are expected to break down, this result
is not to be taken too seriously.

\subsection{Case 2: $\omega\Lambda/\tilde Q\ll 1$
                 and $\omega^2\Lambda/\tilde Q \sim 1$}
\label{case 2} This limit requires that $M_X$ is considerably
smaller then $m_p$ and $\omega$ needs at least to be order 10.
However these assumptions are not unrealistic. For simplicity we
also assume that $\omega\gg \chi+1$, so we can use the results of
section \ref{sparameters}. Furthermore we assume that $|\chi|\leq
\mathcal{O}(1)$, so we can use the approximation
(\ref{logreplacements}) (see also the comment following equation
(\ref{Lren})). Finally we shall drop the $H^4/\tilde{Q}$ terms in
the Friedmann equation, since we will be interested in the regime
where $H$ becomes small. We now summarize all of our assumptions,
\begin{equation} \label{allassumptions}
        H^4/\tilde{Q} \ll H^2 \ll \omega\Lambda \ll \tilde Q
\,,\qquad
        \omega^2\Lambda \sim \tilde Q
\,,\qquad
        \omega\gg \chi+1\qquad |\chi|\leq  1
\,.
\end{equation}
With these assumptions the Friedmann equation (\ref{fried})
becomes
\begin{equation}\label{fried2}
    H^2=\frac{\lambda[t]\tilde Q_0}{3\omega^2}
          \Big[1+(1-3\lambda)3\lambda L-\frac{9}{2}\lambda^2\Big]
\end{equation}
where we have defined
 \begin{equation}\label{definities}
    \begin{split}
    \lambda&\equiv\frac{\omega^2\Lambda[t]}{\tilde Q_0}
               =\frac{\lambda_0}{1-6\lambda_0 t}\\
    \lambda_0&\equiv\frac{\omega^2\Lambda_0}{\tilde Q_0}
                 \simeq \frac{\omega^2}{8\pi}\frac{M_X^4}{m_p^4}
\\
        L&\equiv\ln\Big(\frac{\tilde Q_0\lambda}{\omega H^2}\Big).
\end{split}
\end{equation}
In the second line of (\ref{definities}) we have used the results
of section \ref{sparameters}. \\
If we start the RG flow at $\mu=H$ and let $\mu$ go to zero (see
section~\ref{Renormalization Group}),
there are two possibilities. If $H$ decreases slower
then $\mu$, $\lambda[t]$ increases; otherwise $\lambda[t]$
decreases. Furthermore it is clear that \emph{if} $\lambda$
increases, and $H$ decreases, clearly $L$ will increase. We solve
the equation [{\it r.h.s.} of (\ref{fried2})]=0 in this limit and get
(apart from the trivial solution)
\begin{equation}
    \lambda=\frac{1}{3}+\frac{1}{6}\frac{1}{L}-\frac{1}{4}\frac{1}{L^2}+\mathcal{O}\Big(\frac{1}{L^3}
\end{equation}
Therefore we see that when $H$ goes to zero, $\lambda$ approaches
$1/3$ from above. But this means that $\lambda$ ought to
\emph{decrease}. Fortunately this is exactly what happens: for
$\lambda<\lambda_{\rm crit}$~\footnote{Numerical analysis shows
that $\lambda_{\rm crit}$ is marginally larger then 1/3.}, the
ratio $H/\mu$ grows and thus $\lambda$ increases. However when
$\lambda$ equals $\lambda_{\rm crit}$, $H$ starts to decrease
faster then $\mu$, so from that point onwards, $\lambda$ actually
starts to increase, driving $H^2$ towards zero. Before studying
this process in more detail in section \ref{szop}, let us first
determine how fast $H^2$ goes to zero as $\lambda\searrow 1/3$. To
do this, we rewrite the Friedmann equations (\ref{fried2}) as
follows:
\begin{equation}
    H^2=\frac{\tilde Q_0\lambda}{\omega}
    \exp\bigg[
             \frac{1}{(1-3\lambda)3\lambda}
             \Big(1-\frac{9}{2}\lambda^2
                 -\frac{3\omega^2H^2}{\tilde Q_0\lambda}
             \Big)
         \bigg]
\,.
\label{Friedmann:Q:inverse}
\end{equation}
The limit $H\rightarrow 0$ and $\lambda\searrow 1/3$ from above thus means
approximately:
\begin{equation}\label{exp}
  H^2\simeq \frac{\tilde Q_0}{3\omega}\exp\bigg[-\frac{1}{6(\lambda-1/3)}\bigg]
\,.
\end{equation}
 From this equation we see that, since $\lambda$ depends on
$\ln(H/\mu)$, we actually have approximately that $H\propto\mu$
and thus we get power-law running instead of logarithmic running.
 In fact $H$ goes even faster to zero, since we require that $H$
goes to zero faster then $\mu$ and therefore $\lambda$
'accelerates' towards $1/3$.

\subsubsection{Numerical analysis of the model in the infrared limit
               $\mu\rightarrow 0$}
 \label{szop}

\begin{figure}\label{hversuslambda}
\begin{center}
\includegraphics[width=5.0in]{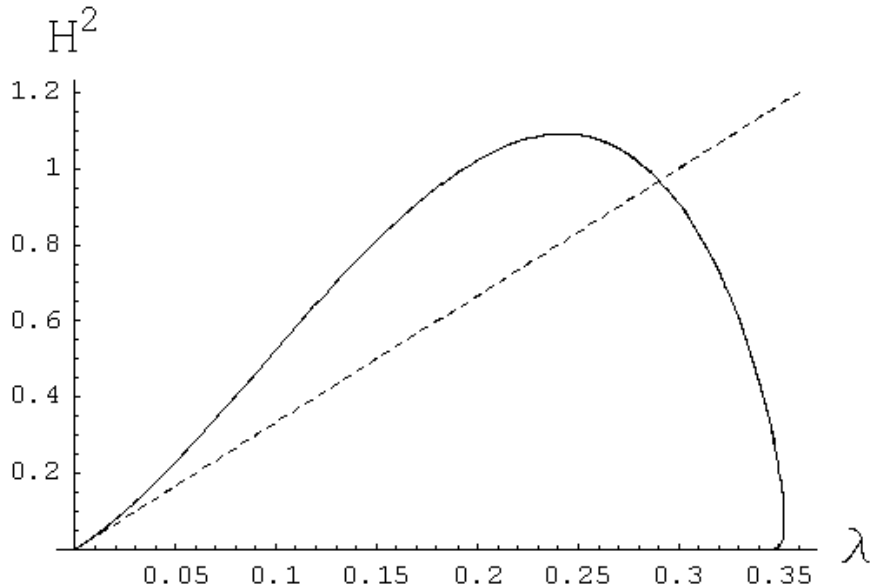}
\caption{$H^2$ versus $\lambda[t]$. The solid curve is the
solution to the Friedmann equation, the dashed line is given by
$H^2=\Lambda[t]/3$.}
\end{center}
\end{figure}

\begin{figure}
\begin{center}
\includegraphics[width=5.0in]{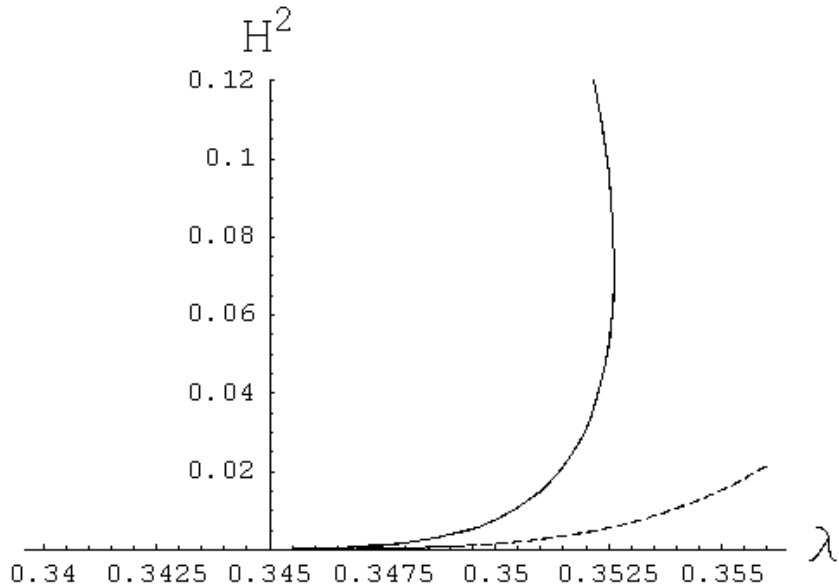}
\caption{Closeup of the region $\lambda[t]>1/3$. The solid curve
is the solution to the Friedman equation. The dashed curve
represents Eq.~(\ref{exp}).}
\label{comparewithexp}
\end{center}
\end{figure}

 In this section we demonstrate that the behavior
described in the previous section, is actually realized in this
model. Unfortunately we are unable to solve the Friedmann equation
analytically for $H$,
 therefore we need to rely on numerical analysis.
For definiteness we choose $\tilde Q_0=1000$ and $\omega =10$,
however none of the qualitative features of the model depends on
these numbers. One does not need to specify $\Lambda_0$, since
$\Lambda_0$ will only define the 'starting point' on the curve
(\ref{hversuslambda}). While we choose two parameters, we actually
have a one parameter family of curves, since the $\tilde{Q}_0$
dependence in (\ref{fried2}) can be removed by defining $h^2\equiv
H^2/ \tilde{Q}_0$.
\\
 We solve the Friedmann equation numerically and plot
$H^2$ versus $\lambda[t]$ (figures \ref{hversuslambda} and
\ref{comparewithexp}).
The first thing we need to consider is whether, if we start with
$0<\lambda_0<1/3$ and let $\mu$ decrease, $\lambda[t]$ increases
or decreases. To see what happens, we consider the Friedmann
equation in the limit $\lambda[t]\rightarrow 0$
\begin{equation}
    H^2\simeq \frac{\lambda[t]\tilde Q_0}{3\omega^2}
\end{equation}
and take the derivative with respect to $\mu$ of this equation to
obtain
\begin{equation}
    \frac{\partial H^2}{\partial
    \mu}=-\Big(1-\frac{Q_0\lambda[t]^2}{H^2\omega^2}\Big)^{-1}\frac{2Q_0\lambda[t]^2}{\mu\omega^2}=-\frac{2Q_0\lambda[t]^2}{\mu\omega^2}+\mathcal{O}(\lambda[t])^4<0
\end{equation}
Since this expression is always negative, we find that upon
decreasing $\mu$, $H$ will increase and therefore $\lambda[t]$
will increase. In other words: starting at the origin of figure
\ref{hversuslambda}, we run along the curve to the right. It is
important to realize that we \emph{cannot} change the running 'direction' on
the curve. Also 'stopping' on the curve is not possible as long as
we keep decreasing $\mu$.
\\
Depending on the value of $\Lambda_0$ (and thus $\lambda_0$)we see
that one actually starts with an effective cosmological constant
that is somewhat larger then one would expect from the unmodified
Friedmann equation (shown by the dashed line in figure
\ref{hversuslambda}).
 Continuing along the curve, at a certain point $H$
starts to decrease. In fact, it is clear that $H$ decreases faster
and faster until a point where it decreases faster then $\mu$ and
therefore eventually $\lambda[t]$ starts to decrease. This
\emph{must} be the case, since this is the only solution to the
Friedmann equation in this regime and we see no good reason why
the Friedmann equation should break down. In figure
\ref{comparewithexp} we show how $H$ goes to zero, compared to our
estimate (\ref{exp}). We see that $H$ decreases even faster then
exponentially as $\lambda[t]$ approaches $\lambda_{\rm crit}=1/3$.
The closer $\lambda[t]$ to 1/3 is, the more accurate (\ref{exp})
becomes.\\
Notice that the fact that $\lambda$ stays finite for all values of
$\mu$ between $H$ and $0$ is an important property of our $t$
parameter (\ref{t}). For other --at first instance sensible--
choices of $t$ like $t=\ln(M_x/\mu)$, $\Lambda$ and therefore
$\lambda$ will hit the Landau pole become infinite.

\subsubsection{Does $H^2$ stay smaller then $\omega\Lambda$?}
\label{sassumpproof}

 In calculating the effective
potential we assumed that (\ref{assumption}) holds, or in the
present case, where $|\chi|\leq 1$:
\begin{equation}\label{assumption2}
    H^2\ll \omega\Lambda.
\end{equation}
We shall now show that this assumption is indeed justified over
the whole range of $\lambda[t]$,
independent of the parameters.\\
First of all notice that the assumptions (\ref{allassumptions})
mean that
\begin{equation} \label{omegaest}
    \omega \geq \mathcal{O}(10).
\end{equation}
Therefore it is clear that \emph{if} the standard Friedmann
equation, $H^2=\Lambda/3$, would be correct, the assumption
(\ref{assumption2}) holds. To see whether this is also true for
our modified Friedmann equation, we take the derivative of
(\ref{fried2}) with respect to $\lambda$
\begin{equation}
    \frac{\partial H^2}{\partial\lambda}
      = \frac{\tilde Q_0\lambda^2}{\omega^2 H^2}(3 \lambda-1)
           \frac{\partial H^2}{\partial\lambda}
        +\frac{\tilde Q_0}{6\omega^2}
            \Big(2+6\lambda-45\lambda^2+6(2-9\lambda)\lambda L\Big),
\end{equation}
with $L$ defined in (\ref{definities}). To calculate at which
$\lambda$ $H$ is maximal, we put $L$ approximately constant and
find
\begin{equation}
    \frac{\partial H^2}{\partial\lambda}=0\quad\rightarrow\quad
    \lambda=\frac{2 L+1+\sqrt{4 L^2+16 L+11}}{3 (6 L+5)}
\end{equation}
and we substitute this in our Friedmann equation, divided by
$\omega\Lambda$ to obtain
\begin{equation}
    \frac{H^2}{\omega\Lambda}\Big|_{\rm max}
         \simeq \frac{(L+1) \left[2 L \left(2 L+\sqrt{4 L (L+4)+11}+12\right)
                  -\sqrt{4 L (L+4)+11}+19\right]}{3 (6 L+5)^2\omega}
\,.
\end{equation}
Since we know that in this regime $H^2=\mathcal{O}(\Lambda)$, it
is safe to say that $L$ grows to as most $\mathcal{O}(10)$. With
this estimate and (\ref{omegaest}) we get approximately
\begin{equation}
    \frac{H^2}{\omega\Lambda}\Big|_{\rm max}\simeq 0.1
\end{equation}
and thus our assumption in (\ref{assumption}) is satisfied.

\subsubsection{The infrared sector}
If the mechanism described above is indeed responsible for the
small effective cosmological constant measured today, we require
that $\mu$ runs at least to a value $\mu_{\rm min}$ such that
$H(\mu_{\rm min})=H_0$. Here $H_0 \simeq 1.5\times 10^{-42}~{\rm
GeV}$ denotes the Hubble parameter as it is measured today. The
analysis presented in this subsection is oversimplified, since we
are neglecting matter and radiation contributions in assuming a
pure de Sitter space-time. We can estimate $\mu_{\rm min}$ by
inverting the relation~(\ref{exp}),
\begin{eqnarray}
 \mu_{\rm min} &=& H_0 \exp
      \left(
        - \frac{4\pi m_p^2}{3\omega^2\Lambda_0}
        + \frac12\frac{1}
           {1+\left[2\ln\Big(\frac{8\pi m_p^2}{3\omega H_0^2}\Big)\right]^{-1}}
     \right)
\nonumber\\
  &\simeq& H_0 \exp
      \left(
        - \frac{4\pi m_p^2}{3\omega^2\Lambda_0}
        + \frac12
     \right).
\label{mu:lowest scale today}
\end{eqnarray}
Notice that the subscript zero for $H$ means 'today', while for
$\Lambda$ it means 'at $\mu=0$.\\
 According to the logic in
section (\ref{Renormalization Group}) the scale~(\ref{mu:lowest
scale today}) corresponds with the largest scale $k$ amplified
during primordial cosmic inflation. Since the number of e-folds is
given by
\begin{equation}
    N_{\rm tot}=\ln\Big(\frac{H_I}{k}\Big),
\end{equation}
with $H_I$ the Hubble parameter during inflation, we get
\begin{eqnarray}
  N_{\rm tot} &\simeq& 60 + \ln\left(\frac{H_0}{\mu_{\rm min}}\right)
\nonumber\\
    &\simeq& 60 + \frac{4\pi m_p^2}{9\omega^2 H_I^2}
\label{N:efolding}
\end{eqnarray}
where we used the usual estimate that
$\ln\Big(\frac{H_I}{H_0}\Big)\simeq 60$ and we assumed that $H_I^2
\simeq \Lambda_0/3$. We can rewrite (\ref{N:efolding}) by noting
that the power spectrum of curvature perturbation has an amplitude
squared,
\begin{equation}
  {\Delta}_{\cal R}^2\simeq \frac{H_I^2}{\pi\epsilon m_p^2}
                       \simeq 2.4\times 10^{-9}
\,, \label{power spectrum}
\end{equation}
where $\epsilon$ is the slow roll parameter in inflation, defined
as $\epsilon = -(dH_I/dt)/H_I^2$ (in scalar inflationary models
its numerical value is typically of the order $10^{-2}$). With
this the total number of e-foldings~(\ref{N:efolding}) becomes,
\begin{equation}
  N_{\rm tot} \simeq 60 + \frac{4}{9\omega^2\epsilon \Delta_{\cal R}^2}
     \simeq 60 + 2\times 10^8\times \frac{1}{\omega^2\epsilon}
\,. \label{N:efolding:2}
\end{equation}
This provides an estimate of the duration of inflation in our
model, which decreases as $1/\omega^2$ as $\omega$ increases. Even
though we are unable to make a direct estimate of the residual
dark energy density in our model, in Eq.~(\ref{N:efolding:2}) we
have provided a link between the the dark energy density (the
Hubble parameter today) and the duration of inflation.

\section{Discussion and conclusion}\label{section discus}

In this paper we consider the one loop vacuum fluctuations in de
Sitter space of a nonsymmetric tensor field, with a mass generated
by a coupling to the cosmological constant and the Ricci scalar.
 While such a field is motivated by nonsymmetric
gravitational theories, the present discussion is independent of
this fact. We believe that qualitatively the results presented here
also hold for {\it e.g.} a scalar field whose mass is generated by $\Lambda$.
 The calculation presented in this work shows that, as one would expect,
the coupling constants of the theory $\Lambda$, $Q= (16\pi G_N)^{-1}$
 and $b$ -- which correspond to the coupling constants of the terms $R^0$, $R$
 and $R^2$ in the gravitational action, respectively -- become renormalized.
 The renormalization depends, of course, on the
renormalization scale $\mu$ and after RG improvement we obtain the
standard result that the constants (in principle) run
logarithmically with $\mu$. Usually this logarithmic running is
used to argue that the RG flow cannot drive the cosmological
constant to zero. Indeed it seems implausible that such a mild
running could explain the smallness of $\Lambda G_N \sim
10^{-122}$, which defines the gravitational hierachy problem.
\\
The main thrust of the present work is the observation that the
observed, effective $\Lambda_{\rm eff}$ may dramatically differ
from the (RG improved) parameter $\Lambda$. The reason is that
$\Lambda_{\rm eff}$ is given by the Friedmann equation,
$H^2=\Lambda_{\rm eff}/3$, and this equation gets modified by the
loop effects. This modification is typically of the order
$m^2/m_p^2$, where $m$ is the mass of the field whose contribution
we are calculating. This quantity will therefore be in general
suppressed. However in our case $m\sim \Lambda$ and standard
quantum field theory calculations tell us that $\Lambda$ may be
comparable to $m_p^2$. In other words, a large $\Lambda$ implies
substantial modification of the Friedmann equation by loop quantum
corrections. At the same time our analysis shows that the Newton
constant remains to a good approximation frozen. This is
important, since a fast running of the Newton constant could be
disastrous in the view of Solar system tests of GR.


 Furthermore, we find that as $\mu$ runs towards zero,
$H$ -- and thus $\Lambda_{\rm eff}$ -- changes mildly for some
range of $\mu$. However from a certain value of $\mu$, $H$ starts
to decrase rapidly, eventually even faster then linearly,
$H\propto \mu$. Therefore, while $\Lambda$ changes only
logarithmically with $\mu$, $\Lambda_{\rm eff}$ can decrease as
power law. In this manner our model circumvents the usual
statement that RG flow cannot relax the cosmological constant. It
is important to stress that this is independent of the size of the
initial $\Lambda$. The initial conditions influence only at what
value of $\mu$ the dramatic decrease in $\Lambda_{\rm eff}$ begins
to take place.
\\
An important point we wish to address now is why do we sent $\mu$
to zero? To understand this, it is first of all necessary to
understand that the starting point of our running, $\mu=H$, is in
the ultraviolet, where both $\Lambda$ and $H$ are of the order the
Planck scale $m_p$. However in de Sitter space the Hubble
parameter is not the lowest possible scale. In fact, many physical
modes have wavelengths, that during an inflationary de Sitter
phase get amplified and their wavelength grows larger then the
hubble length, $H^{-1}$. In order to get a theory that takes
proper account of these super-Hubble modes, the infrared $\mu$
should be taken as small as the smallest energy of these modes.
Since we are interested in the behavior of the theory on large
(cosmological) scales, this is exactly what one should do.
\\
Another interesting question is how applicable our model is to the
real world. Strictly speaking, our results are only valid in de
Sitter space and one could wonder if our results also apply to, for
example, quasi-de Sitter spaces where $H$ is a (mild) function of
time. Quasi-de Sitter spaces are relevant for inflationary models
and also in the case whenever $\Lambda_{\rm eff}$
dominates the energy density of the Universe, as it is today.
Even though we have performed our analysis in de Sitter space,
we expect that our analysis should apply whenever $dH/dt \ll H^2$.
This condition is satisfied if $\Lambda_{\rm eff}/(8\pi G_N)$
dominates the energy density of the Universe, which is also
(marginally) true today.

 There might be implications of our model for inflation. One could envisage
a scenario, where inflation is caused by a large cosmological
constant and as times goes on more and more of infrared modes are produced,
pushing the relevant RG scale $\mu$ further and further to the infrared,
such that $\Lambda_{\rm eff}$ eventually decreases, terminating
inflation, resulting in a standard decelerating Friedmann universe.
This type of reasoning has also been pursued in
Refs.~\cite{Tsamis:1996qq,Tsamis:1996qm,Abramo:1997hu}
by using different techniques.
An explicit calculation is necessary in the present or similar model
however in order to illuminate the questions concerning inflationary
dynamics and cosmological perturbations.
 Based on the assumption that the scale $\mu$ corresponds to the
largest scale amplified during inflation,
in section~\ref{case 2} we derive a relation between
the total number of e-foldings during inflation in our model (driven by
a bare cosmological constant) and the dark energy density as measured today.
However at the moment we are unable to provide a direct estimate
of the dark energy density.\\
While our model shows interesting behavior, there are some
potential problems. First of all one might argue that our model
suggests that at small scales $\Lambda_{\rm eff}$ is huge, and
this could present problems to e.g. solar system measurements.
However, this reasoning is not correct. Even in small scale
experiments, the large scale modes still have their effect on the
vacuum. In other words: it is impossible to decouple the infrared
modes from the vacuum.
\\
a more serious concern is the validity of the perturbative
approach. Indeed in our model tree level and one loop
contributions are comparable and therefore it is not clear whether
our results hold if one includes higher loop corrections.
Furthermore there is a similar problem in the renormalization
procedure. The choice of the renormalization scheme and the
running parameter $t$ are in principle arbitrary. Different
choices however differ only at higher loop order. Since higher
loop contributions are probably substantial in our model, there is
no way of telling which effects are induced by our
choices and which effects are really physical.\\
Despite these problems we do believe that the following is true. A
field whose mass is generated by the cosmological constant might
generate a huge back-reaction on the vacuum. This back-reaction
can be of the same order of magnitude as the original cosmological
constant. It might be that the exact structure of the modified
Friedmann equation is such that, because of the RG running of
$\Lambda$, the effective cosmological constant becomes tiny,
without fine tuning and regardless of the initial condition. In
this work we have shown an explicit example of this behavior.

\section{acknowledgement}

 We would like to thank Frank Saueressig for sharing his
insights in the renormalization group.

\appendix
\section{Separating the finite and infinite parts of $\Gamma_B$}
\label{appendix effective action} To separate the infinite from
the finite part of the effective action $\Gamma$~(\ref{gammab}), we use the
following expression
\begin{equation}
\frac{\Gamma(\frac{D-1}{2}+\nu_n)\Gamma(\frac{D-1}{2}-\nu_n)}
     {\Gamma(\frac{1}{2}+\nu_n)\Gamma(\frac{1}{2}-\nu_n)}
 = \bigg(\Big(\frac{D-3}{2}\Big)^2-\nu_n^2\bigg)
   \bigg[1+\frac{D-4}{2}\Big(\psi(\frac{1}{2}+\tilde{\nu}_n)+\psi(\frac{1}{2}
          -\tilde{\nu}_n)\Big)
    \bigg]+\mathcal{O}\left((D-4)^2\right)
\,, \label{Gamma:decomp}
\end{equation}
where $\tilde \nu_n = \nu_n|_{D=4}$ and $\psi(z)$ is the digamma
function, defined by
\begin{equation}
    \psi(z)\equiv\frac{\Gamma'(z)}{\Gamma(z)}
\,.
\end{equation}

\subsection{Evaluating $\Gamma_w$}

 By making use of Eqs.~(\ref{gammab}) and~(\ref{Gamma:decomp}) we can break
$\Gamma_w$ into the infinite and finite parts as follows:
\begin{equation}
    \begin{split}
        \Gamma_w&=-\int_x\sqrt{-g}\Bigg[\int
        dw\frac{D-1}{4(4\pi)^{D/2}}H^D\Gamma(1-D/2)\Bigg((D-2)(D-4+w)-(2(D-5)+w)\Bigg)\\
        &\qquad+\frac{3H^4}{64\pi^2}\int dw
         \Bigg(2w\Big(\psi(1/2+\tilde{\nu}_2)+\psi(1/2-\tilde{\nu}_2)\Big)
   -(w-2)\Big(\psi(1/2+\tilde{\nu}_3)+\psi(1/2-\tilde{\nu}_3)\Big)\Bigg)\Bigg]
\,.
    \end{split}
\nonumber
\end{equation}
This breakdown is not unique (since we can always shift a finite
part of the infinite part back and forth), but when taken together
with the counterterms, this nonuniqueness has no physical
consequence. The integral in the infinite part is easy to
evaluate:
\begin{equation}\label{Gwinf}
\begin{split}
      \Gamma_{w-inf}
   =-\int_x\sqrt{-g}\Bigg[
                         \frac{D-1}{4(4\pi)^{D/2}}
                  \Gamma(1-D/2)\Bigg(&\frac{1}{2}(D-3)(\omega\Lambda)^2H^{D-4}
\\
      &+(D^2-8D+18+(D-3)\chi)\omega\Lambda H^{D-2}\\
      &+\Big((D^2-8D+18)\chi+\frac{1}{2}(D-3)\chi^2\Big)H^D\Bigg)\Bigg]
\,,
    \end{split}
\end{equation}
where we used that~(\ref{m2_Fabc})
\begin{equation}
    m^2=\omega\Lambda+\chi H^2
\,.
\end{equation}
To evaluate the finite part we assume that
\begin{equation}
    w=\frac{\omega\Lambda+\chi H^2}{H^2}\gg 1,
\end{equation}
which we justify in section \ref{sassumpproof}. With this
assumption we can use
\begin{equation}
    \begin{split}
        &\psi(1/2+\nu_2)+\psi(1/2-\nu_2)
   =\ln(w)-\frac{1}{3w}-\frac{1}{15w^2}+\mathcal{O}(w^{-3})
\\
        &\psi(1/2+\nu_3)+\psi(1/2-\nu_3)
  =\ln(w)-\frac{7}{3w}-\frac{41}{15w^2}+\mathcal{O}(w^{-3})
    \end{split}
\end{equation}
and find:
\begin{equation}
    \begin{split} \label{Gwfin}
        \Gamma_{w-fin}
&=-\int_x\sqrt{-g}\frac{3}{64\pi^2}H^4
 \Bigg[-\frac{31}{15}\ln(w)
       +\Big(-\frac{1}{3}+2\ln(w)\Big)w
       +\Big(-\frac{1}{4}+\frac{1}{2}\ln(w)\Big)w^2
 \Bigg]
\\
        &=-\int_x\sqrt{-g}\frac{3}{63\pi^2}
  \Bigg[-\frac{31}{15}H^4\ln(w)
       +\Bigg(\frac{(\omega\Lambda)^2 }{2}
             +H^2\omega\Lambda (2+\chi)
             +H^4\chi\Big(2+\frac{\chi}{2}\Big)
          \Bigg)\ln(w)
\\
        &\qquad\qquad\qquad\qquad\quad
    -\frac{(\omega\Lambda)^2 }{4}-H^2\omega\Lambda
     \Big(\frac{1}{3}+\frac{\chi}{2}\Big)
     -H^4\chi\Big(\frac{1}{3}+\frac{\chi}{4}\Big)
  \Bigg]
\,.
    \end{split}
\end{equation}

\subsection{Evaluating $\Gamma_H$}

 From Eq.~(\ref{gammab}) we can write
\begin{equation}
\label{Gamma_H}
    \begin{split}
    \Gamma_H=&-\int_x\sqrt{-g}
  \Bigg[
\\
    &\int dH^2(D-1)\frac{H^{D-2}}{4(4\pi)^{D/2}}
           \Gamma(1-D/2)\Bigg((D-2)(D-4+w)w+(D-6-w)(2(D-5)+w)\Bigg)
\\
    +&\frac{3}{64\pi^2}\int dH^2H^{D-2}
          \Bigg(2w^2\Big(\psi(1/2+\tilde{\nu}_2)+\psi(1/2-\tilde{\nu}_2)\Big)
          -(w^2-4)\Big(\psi(1/2+\tilde{\nu}_3)+\psi(1/2-\tilde{\nu}_3)\Big)
          \Bigg)
  \Bigg]\,.
    \end{split}
\nonumber
\end{equation}
We start with the infinite part:
\begin{equation}\label{GHinf}
    \begin{split}
         \Gamma_{H-inf}&=-\int_x\sqrt{-g}
         \Bigg[\frac{D-1}{4(4\pi)^{D/2}}\frac{2}{D}H^D\Gamma(1-D/2)
           \bigg(2(D-5)(D-6)+(D-4)(D-3)w+(D-3)w^2\bigg)
         \Bigg]
\\
            &=-\int_x\sqrt{-g}
      \Bigg[\frac{D-1}{4(4\pi)^{D/2}}\frac{2}{D}
          \bigg(H^D\Big(2(D-5)(D-6)+(D-4)(D-3)\chi+(D-3)\chi^2\Big)
\\
            &\qquad\qquad\qquad+H^{D-2}\omega\Lambda
             \Big((D-4)(D-3)+2(D-3)\chi\Big)
             +H^{D-4}(\omega\Lambda)^2 (D-3)\bigg)
       \Bigg]\,.
    \end{split}
\end{equation}
The finite part is easily evaluated and gives
\begin{equation}\label{GHfin}
    \begin{split}
        \Gamma_{H-fin}&=-\int_x\sqrt{-g}\frac{3}{64\pi^2}H^4
               \bigg(2\ln(w)+\frac{w^2}{2}\ln(w)+\frac{5}{6}w+\frac{13}{10}
               \bigg)\\
        &\hskip -1cm
            =-\int_x\sqrt{-g}\frac{3}{64\pi^2}
               \bigg(2H^4\ln(w)
                +\Big(\frac{(\omega\Lambda)^2 }{2}+\chi\omega\Lambda H^2
                       +\frac12\chi^2H^4
                  \Big)\ln(w)
               +\frac{13}{10}H^4+\frac{5}{6}(\chi H^4+\omega\Lambda H^2)
              \bigg)
\,.
    \end{split}
\end{equation}

\subsection{Combining the results}
Since in $D=4$ our counterterms are of the form $R^0$,  $R^1$,
 $R^2...$, we need to expand the $D$ dependent powers in $H$. To
achieve this, one needs to introduce an arbitrary mass scale
$\mu$, as can be seen from the following identities:
\begin{equation} \label{introducing mu}
    \begin{split}
        H^D&=\mu^{D-4}\bigg(H^4+\frac{D-4}{2}H^4\ln\Big(\frac{H^2}{\mu^2}\Big)\bigg)
             + {\cal O}\Big((D-4)^2\Big)
\\
        H^{D-2}\omega\Lambda &=\mu^{D-4}\bigg(H^2\omega\Lambda
              +\frac{D-4}{2}H^2\omega\Lambda
              \ln\Big(\frac{H^2}{\mu^2}\Big)\bigg)
              + {\cal O}\Big((D-4)^2\Big)
\\
        H^{D-4}(\omega\Lambda)^2 &=\mu^{D-4}\bigg((\omega\Lambda)^2
               +\frac{D-4}{2}(\omega\Lambda)^2
               \ln\Big(\frac{H^2}{\mu^2}\Big)\bigg)
               + {\cal O}\Big((D-4)^2\Big).
    \end{split}
\end{equation}
Notice that this statement is equivalent to the argument used in
e.g.~\cite{Itzykson:1980rh}, where $\mu$ is introduced in order to
give physical constants the right dimensionality. Using
(\ref{introducing mu}), we combine Eqs.~(\ref{Gwinf}),
(\ref{Gwfin}), (\ref{GHinf}) and (\ref{GHfin}) to obtain our final
result for the 1-loop effective action for the $B$-field
\begin{equation}
   \begin{split}
        \Gamma_B=&-\int_x\sqrt{-g}\Bigg[\\
        &\frac{D-1}{4(4\pi)^{D/2}}\Gamma(1-D/2)
     \Bigg(H^4\frac{1}{2D}
          \Big[8(D-5)(D-6)+2(24+D(D^2-6D+4))\chi+(D-3)(D+4)\chi^2
          \Big]
      \\
        &\qquad+H^2\omega\Lambda \Big(4 + D^2+D(\chi-6) -
        \frac{12(\chi-2)}{D}+\chi\Big)+(\omega\Lambda)^2 \frac{(D-3)(D+4)}{2D}\Bigg)\mu^{D-4}\\
       &+\frac{3}{64\pi^2}\Bigg(\Big((\omega\Lambda)^2 +2(1+\chi)\omega\Lambda H^2+2(1+\chi+\frac{\chi^2}{2})H^4\Big)\ln\Big(\frac{\omega\Lambda +\chi H^2}{\mu^2}\Big)-\frac{31}{15}H^4\ln\Big(\frac{\omega\Lambda +\chi H^2}{H^2}\Big)\\
       &\qquad-\frac{(\omega\Lambda)^2 }{4}+\frac{H^2\omega\Lambda }{2}(1-\chi)+\frac{\chi
       H^4}{2}(1-\frac{\chi}{2})+\frac{13}{10}H^4\Bigg)\Bigg]+\mathcal{O}\Big(H^4\frac{H^2}{m^2}\Big)\\
       \equiv&\int_x\delta\mathcal{L}
   \end{split}
\end{equation}

\section{Method of characteristics}\label{ap method of char}
In this Appendix we shall describe how the method of
characteristics works~\cite{Kleinert:2001ax}\cite{Faux}. The
requirement that physics is independent of $\mu$, is equivalent to
requiring that the theory is invariant under the transformation
\begin{equation} \label{transformation1}
    \mu\rightarrow \mu {\rm e}^t\equiv\bar\mu[t],
\end{equation}
with $t$ a real number. Since our coupling parameters $\Lambda$,
$Q$ and $b$ depend on $\mu$, the transformation
(\ref{transformation1}) means that
\begin{equation}\label{transformation2}
    \begin{split}
        \Lambda&\rightarrow \bar{\Lambda}[t]\\
        Q&\rightarrow \bar{Q}[t]\\
        b&\rightarrow \bar{b}[t].
    \end{split}
\end{equation}
Quantities that we require to be independent of $\mu$, like the
effective action, need to be independent of this transformation
and therefore
\begin{equation}
    \Gamma(\Lambda,Q,\mu)=\Gamma(\bar{\Lambda}[t],\bar{Q}[t],\bar{b}[t],\bar{\mu}[t]).
\end{equation}
This can only be true if
\begin{equation}
\begin{split}
    0&=\frac{d}{dt}\Gamma(\bar{\Lambda}[t],\bar{Q}[t],\bar{b}[t],\bar{\mu}[t])\\
    &=\Big(\bar{\mu}\frac{\partial}{\partial\bar{\mu}}
    +\frac{d\bar{\Lambda}}{d t}\frac{\partial}{\partial \bar{\Lambda}}
    +\frac{d\bar{Q}}{dt}\frac{\partial}{\partial \bar{Q}}
     +\frac{d\bar{b}}{d t}\frac{\partial}{\partial \bar{b}}\Big)
    \Gamma(\bar{\Lambda}[t],\bar{Q}[t],\bar{b}[t],\bar{\mu}[t])
\,.
\end{split}
\end{equation}
Comparing this with the Callan-Symanzik equation
\begin{equation}
    \Big(\mu\frac{\partial}{\partial\mu}
    +\beta_\Lambda\frac{\partial}{\partial \Lambda}
    +\beta_Q\frac{\partial}{\partial Q}
     +\beta_b\frac{\partial}{\partial \bar{b}}\Big)
    \Gamma(\Lambda,Q,b,\mu)=0,
\end{equation}
we conclude that the transformation (\ref{transformation1}) and
(\ref{transformation2}), with parameters given by
\begin{equation}
\begin{split}
\frac{d\bar{\Lambda}[t]}{d t}
       &= \beta_\Lambda[\bar{\Lambda}[t],\bar{Q}[t]]
\qquad
     \bar{\Lambda}[0] = \Lambda
\\
\frac{d\bar{Q}[t]}{d t}&=\beta_Q[\bar{\Lambda}[t]]\qquad\qquad\bar{Q}[0]=Q
\\
\frac{d\bar{b}[t]}{d t}&=\beta_b
\qquad\qquad\qquad\;\,\,\,
      \bar{b}[0] = b
\end{split}
\end{equation}
enforces the Callan-Symanzik equation for the effective action $\Gamma$.

\newpage
\bibliographystyle{utcaps}
\bibliography{LambdaAsym}

\providecommand{\href}[2]{#2}\begingroup\raggedright\begin{thebibliography}{10}

\bibitem{Weinberg:1988cp}
S.~Weinberg, ``The cosmological constant problem,'' {\em Rev. Mod. Phys.} {\bf
  61} (1989)
1--23.

\bibitem{Nobbenhuis:2006yf}
S.~Nobbenhuis, ``The cosmological constant problem, an inspiration for new
  physics,''
\href{http://arXiv.org/abs/gr-qc/0609011}{{\tt gr-qc/0609011}}.

\bibitem{Lamoreaux:1996wh}
S.~K. Lamoreaux, ``Demonstration of the Casimir force in the 0.6 to 6
  micrometers range,'' {\em Phys. Rev. Lett.} {\bf 78} (1997)
5--7.

\bibitem{Casimir:1948dh}
H.~B.~G. Casimir, ``On the attraction between two perfectly conducting
  plates,'' {\em Kon. Ned. Akad. Wetensch. Proc.} {\bf 51} (1948)
793--795.

\bibitem{Jaffe:2005vp}
R.~L. Jaffe, ``The Casimir effect and the quantum vacuum,'' {\em Phys. Rev.}
  {\bf D72} (2005) 021301,
\href{http://arXiv.org/abs/hep-th/0503158}{{\tt hep-th/0503158}}.

\bibitem{Perlmutter:1998np}
{\bf Supernova Cosmology Project} Collaboration, S.~Perlmutter {\em et al.},
  ``Measurements of Omega and Lambda from 42 High-Redshift Supernovae,'' {\em
  Astrophys. J.} {\bf 517} (1999) 565--586,
\href{http://arXiv.org/abs/astro-ph/9812133}{{\tt astro-ph/9812133}}.

\bibitem{Riess:1998cb}
{\bf Supernova Search Team} Collaboration, A.~G. Riess {\em et al.},
  ``Observational Evidence from Supernovae for an Accelerating Universe and a
  Cosmological Constant,'' {\em Astron. J.} {\bf 116} (1998) 1009--1038,
\href{http://arXiv.org/abs/astro-ph/9805201}{{\tt astro-ph/9805201}}.

\bibitem{Astier:2005qq}
P.~Astier {\em et al.}, ``The Supernova Legacy Survey: Measurement of
  $Omega_M$, $Omega_Lambda$ and w from the First Year Data Set,'' {\em Astron.
  Astrophys.} {\bf 447} (2006) 31--48,
\href{http://arXiv.org/abs/astro-ph/0510447}{{\tt astro-ph/0510447}}.

\bibitem{Riess:2004nr}
{\bf Supernova Search Team} Collaboration, A.~G. Riess {\em et al.}, ``Type Ia
  Supernova Discoveries at z>1 From the Hubble Space Telescope: Evidence for
  Past Deceleration and Constraints on Dark Energy Evolution,'' {\em Astrophys.
  J.} {\bf 607} (2004) 665--687,
\href{http://arXiv.org/abs/astro-ph/0402512}{{\tt astro-ph/0402512}}.

\bibitem{Spergel:2006hy}
D.~N. Spergel {\em et al.}, ``Wilkinson Microwave Anisotropy Probe (WMAP) three
  year results: Implications for cosmology,''
\href{http://arXiv.org/abs/astro-ph/0603449}{{\tt astro-ph/0603449}}.

\bibitem{Peskin:1995ev}
M.~E. Peskin and D.~V. Schroeder, ``An Introduction to quantum field theory,''.
  Reading, USA: Addison-Wesley (1995) 842 p.

\bibitem{Itzykson:1980rh}
C.~Itzykson and J.~B. Zuber, ``Quantum field theory,''. New York, Usa:
  Mcgraw-hill (1980) 705 P.(International Series In Pure and Applied Physics).

\bibitem{Shapiro:1999zt}
I.~L. Shapiro and J.~Sola, ``On the scaling behavior of the cosmological
  constant and the possible existence of new forces and new light degrees of
  freedom,'' {\em Phys. Lett.} {\bf B475} (2000) 236--246,
\href{http://arXiv.org/abs/hep-ph/9910462}{{\tt hep-ph/9910462}}.

\bibitem{Shapiro:2000dz}
I.~L. Shapiro and J.~Sola, ``Scaling behavior of the cosmological constant:
  Interface between quantum field theory and cosmology,'' {\em JHEP} {\bf 02}
  (2002) 006,
\href{http://arXiv.org/abs/hep-th/0012227}{{\tt hep-th/0012227}}.

\bibitem{Shapiro:2006qx}
I.~L. Shapiro and J.~Sola, ``Cosmological constant problems and renormalization
  group,''
\href{http://arXiv.org/abs/gr-qc/0611055}{{\tt gr-qc/0611055}}.

\bibitem{Reuter:1996cp}
M.~Reuter, ``Nonperturbative Evolution Equation for Quantum Gravity,'' {\em
  Phys. Rev.} {\bf D57} (1998) 971--985,
\href{http://arXiv.org/abs/hep-th/9605030}{{\tt hep-th/9605030}}.

\bibitem{Reuter:2001ag}
M.~Reuter and F.~Saueressig, ``Renormalization group flow of quantum gravity in
  the Einstein-Hilbert truncation,'' {\em Phys. Rev.} {\bf D65} (2002) 065016,
\href{http://arXiv.org/abs/hep-th/0110054}{{\tt hep-th/0110054}}.

\bibitem{Reuter:2004nx}
M.~Reuter and H.~Weyer, ``Quantum gravity at astrophysical distances?,'' {\em
  JCAP} {\bf 0412} (2004) 001,
\href{http://arXiv.org/abs/hep-th/0410119}{{\tt hep-th/0410119}}.

\bibitem{Prokopec:2006yh}
T.~Prokopec, ``A solution to the cosmological constant problem,''
\href{http://arXiv.org/abs/gr-qc/0603088}{{\tt gr-qc/0603088}}.

\bibitem{Moffat:1994hv}
J.~W. Moffat, ``Nonsymmetric gravitational theory,'' {\em Phys. Lett.} {\bf
  B355} (1995) 447--452,
\href{http://arXiv.org/abs/gr-qc/9411006}{{\tt gr-qc/9411006}}.

\bibitem{Clayton:1995yi}
M.~A. Clayton, ``Massive NGT and spherically symmetric systems,'' {\em J. Math.
  Phys.} {\bf 37} (1996) 395--420,
\href{http://arXiv.org/abs/gr-qc/9505005}{{\tt gr-qc/9505005}}.

\bibitem{Janssen:2006jx}
T.~Janssen and T.~Prokopec, ``Problems and hopes in nonsymmetric gravity,''
\href{http://arXiv.org/abs/gr-qc/0611005}{{\tt gr-qc/0611005}}.

\bibitem{Mao:2006bb}
Y.~Mao, M.~Tegmark, A.~Guth, and S.~Cabi, ``Constraining Torsion with Gravity
  Probe B,''
\href{http://arXiv.org/abs/gr-qc/0608121}{{\tt gr-qc/0608121}}.

\bibitem{Prokopec:2006kr}
T.~Prokopec and W.~Valkenburg, ``Antisymmetric Metric Field as Dark Matter,''
\href{http://arXiv.org/abs/astro-ph/0606315}{{\tt astro-ph/0606315}}.

\bibitem{Prokopec:2005fb}
T.~Prokopec and W.~Valkenburg, ``The cosmology of the nonsymmetric theory of
  gravitation,'' {\em Phys. Lett.} {\bf B636} (2006) 1--4,
\href{http://arXiv.org/abs/astro-ph/0503289}{{\tt astro-ph/0503289}}.

\bibitem{Valkenburg}
W.~Valkenburg, ``Linearized nonsymmetric metric pertubations in cosmology,''
  {\em master's thesis at the ITP of Utecht University, Available at
  http://www1.phys.uu.nl/wwwitf/teaching/thesis.htm} (2006).

\bibitem{Janssen:2006nn}
T.~Janssen and T.~Prokopec, ``Instabilities in the nonsymmetric theory of
  gravitation,'' {\em Class. Quant. Grav.} {\bf 23} (2006) 4967--4982,
\href{http://arXiv.org/abs/gr-qc/0604094}{{\tt gr-qc/0604094}}.

\bibitem{Kahya:2005kj}
E.~O. Kahya and R.~P. Woodard, ``Charged scalar self-mass during inflation,''
  {\em Phys. Rev.} {\bf D72} (2005) 104001,
\href{http://arXiv.org/abs/gr-qc/0508015}{{\tt gr-qc/0508015}}.

\bibitem{Tsamis:2006gj}
N.~C. Tsamis and R.~P. Woodard, ``A maximally symmetric vector propagator,''
\href{http://arXiv.org/abs/gr-qc/0608069}{{\tt gr-qc/0608069}}.

\bibitem{Spradlin:2001pw}
M.~Spradlin, A.~Strominger, and A.~Volovich, ``Les Houches lectures on de
  Sitter space,''
\href{http://arXiv.org/abs/hep-th/0110007}{{\tt hep-th/0110007}}.

\bibitem{Allen:1985ux}
B.~Allen, ``Vacuum states in de Sitter space,'' {\em Phys. Rev.} {\bf D32}
  (1985)
3136.

\bibitem{Allen:1987tz}
B.~Allen and A.~Folacci, ``The massless minimally coupled scalar field in de
  Sitter space,'' {\em Phys. Rev.} {\bf D35} (1987)
3771.

\bibitem{Chernikov:1968zm}
N.~A. Chernikov and E.~A. Tagirov, ``Quantum theory of scalar fields in de
  Sitter space-time,'' {\em Annales Poincare Phys. Theor.} {\bf A9} (1968)
109.

\bibitem{Tsamis:2005je}
N.~C. Tsamis and R.~P. Woodard, ``Dimensionally regulated graviton 1-point
  function in de Sitter,'' {\em Annals Phys.} {\bf 321} (2006) 875--893,
\href{http://arXiv.org/abs/gr-qc/0506056}{{\tt gr-qc/0506056}}.

\bibitem{Birrell:1982ix}
N.~D. Birrell and P.~C.~W. Davies, ``Quantum fields in curved space,''.
  Cambridge, Uk: Univ. Pr. ( 1982) 340p.

\bibitem{Coleman:1973jx}
S.~R. Coleman and E.~Weinberg, ``Radiative corrections as the origin of
  spontaneous symmetry breaking,'' {\em Phys. Rev.} {\bf D7} (1973)
1888--1910.

\bibitem{Kleinert:2001ax}
H.~Kleinert and V.~Schulte-Frohlinde, ``Critical properties of
  phi**4-theories,''. River Edge, USA: World Scientific (2001) 489 p.

\bibitem{Faux}
M.~Faux, ``Notes on renormalization,''. Available at
  http://www.math.columbia.edu/~faux/geometry/renormalization.ps.

\bibitem{Tsamis:1996qq}
N.~C. Tsamis and R.~P. Woodard, ``Quantum Gravity Slows Inflation,'' {\em Nucl.
  Phys.} {\bf B474} (1996) 235--248,
\href{http://arXiv.org/abs/hep-ph/9602315}{{\tt hep-ph/9602315}}.

\bibitem{Tsamis:1996qm}
N.~C. Tsamis and R.~P. Woodard, ``The quantum gravitational back-reaction on
  inflation,'' {\em Annals Phys.} {\bf 253} (1997) 1--54,
\href{http://arXiv.org/abs/hep-ph/9602316}{{\tt hep-ph/9602316}}.

\bibitem{Abramo:1997hu}
L.~R.~W. Abramo, R.~H. Brandenberger, and V.~F. Mukhanov, ``The energy-momentum
  tensor for cosmological perturbations,'' {\em Phys. Rev.} {\bf D56} (1997)
  3248--3257,
\href{http://arXiv.org/abs/gr-qc/9704037}{{\tt gr-qc/9704037}}.

\bibitem{Mottola:1984ar}
E.~Mottola, ``Particle creation in de Sitter space,'' {\em Phys. Rev.} {\bf
  D31} (1985)
754.

\bibitem{Embacher:1986pt}
H.~G. Embacher, G.~Grubl, and R.~Patek, ``Gauge invariant energy momentum
  tensor for massive QED,'' {\em Phys. Rev.} {\bf D33} (1986)
1162--1165.

\bibitem{Ruegg:2003ps}
H.~Ruegg and M.~Ruiz-Altaba, ``The Stueckelberg field,'' {\em Int. J. Mod.
  Phys.} {\bf A19} (2004) 3265--3348,
\href{http://arXiv.org/abs/hep-th/0304245}{{\tt hep-th/0304245}}.

\end{thebibliography}\endgroup
\newpage

\end{document}